\documentclass[pre,twocolumn,showpacs]{revtex4}

\usepackage{epsfig,enumerate,graphicx}
\usepackage{verbatim}

\newcommand{\Ai}[1]{#1}

\newcommand{\Om}{\Omega}
\newcommand{\om}{\omega}
\newcommand{\tbf}{\textbf}

\newcommand{\req}[1]{(\ref{#1})}

\newcommand{\sche}{\Ai{Scheidegger}}

\newcommand{\toku}{\Ai{Tokunaga}}

\newcommand{\etal}{et al.}

\begin{document}

\title{Unified View of Scaling Laws for River Networks \\
\normalsize{Phys.\ Rev.\ E \textbf{59}(5), May 1999}}

\author{Peter Sheridan Dodds}
\thanks{Author to whom correspondence should be addressed}
\email{dodds@segovia.mit.edu}
\homepage{http://segovia.mit.edu/}
\affiliation{Department of Mathematics
and Department  of Earth, 
Atmospheric and Planetary Sciences,
Massachusetts Institute of Technology,
Cambridge, MA 02139.}

\author{Daniel H. Rothman}
\email{dan@segovia.mit.edu}
\affiliation{Department  of Earth, 
Atmospheric and Planetary Sciences,
Massachusetts Institute of Technology, 
Cambridge, MA 02139.}

\date{\today}

\begin{abstract}
Scaling laws that describe the
structure of river networks
are shown to follow from three simple assumptions.
These assumptions are: (1) river networks are structurally self-similar,
(2) single channels are self-affine, and (3)
overland flow into channels occurs over a characteristic distance
(drainage density is uniform).
We obtain a complete set of scaling relations connecting 
the exponents of these scaling laws and 
find that only two of these exponents are independent.
We further demonstrate that the two predominant descriptions of
network structure (Tokunaga's law and Horton's laws) are
equivalent in
the case of landscapes with uniform drainage density.
The results are tested with data from both real landscapes and
a special class of random networks.
\end{abstract}
\pacs{92.40.Fb, 92.40.Gc, 68.70.+w, 64.60.Ht}

\maketitle

\section{Introduction}
If it is true that scaling laws abound in nature~\cite{mandelbrot83},
then river networks stand as a superb epitome of this phenomenon.
For over half a century, researchers have uncovered numerous
power laws and scaling behaviors
in the mathematical description of river networks~\cite{horton45,langbein47,strahler52,hack57,tarboton88,labarbera89,tarboton90,maritan96a}.
These scaling laws, which are usually parameterized by exponents 
or ratios of fundamental quantities,
have been used to validate 
scores of numerical and theoretical models of landscape 
evolution~\cite{leopold62,shreve66,howard71b,stark91,meakin91,willgoose91a,willgoose91c,willgoose91,kramer92,leheny93,sun94,sun94b,somfai97,rodriguez-iturbe97}
and have even been invoked as evidence of self-organized
criticality~\cite{rodriguez-iturbe97,bak97}.
However, despite this widespread usage, there is as yet 
no fundamental understanding of
the origin of scaling laws in river networks.

It is the principal aim of this paper to bring together
a large family of these scaling laws
within a simple, logical framework.
In particular, we demonstrate that from a base of
three assumptions regarding network geometry,
all scaling laws involving planform quantities may be obtained.
The worth of these consequent scaling laws is
then seen to rest squarely upon the shoulders of the
structural assumptions themselves.  We also
simplify the relations between the derived
laws, demonstrating that only two scaling exponents are
independent.

The paper is composed in the following manner.
We first present preliminary 
definitions of network quantities and a list of
empirically observed scaling laws.
Our assumptions will next be fully stated along with evidence for
their validity.
Several sections will then detail the derivations of the various
scaling laws, being a combination of both new insights of our own
as well as previous results.   
Progressing in a systematic way
from our assumptions, we will also be required to amend
several inconsistencies persistent in other analyses.
The theory will be tested with comparisons to
data taken from real landscapes
and \Ai{Scheidegger}'s 
random network model~\cite{scheidegger67b,scheidegger90}.

\section{The ordering of streams}
A basic tool used in the analysis of river networks is the
device of stream ordering.  
A stream ordering is any scheme that
attaches levels of significance to streams throughout
a basin.  
Most orderings identify the smallest tributaries
as lowest order streams and the main or `trunk' stream as being of
highest order with the intermediary `stream segments' spanning this
range in some systematic fashion.  
Stream orderings allow for
logical comparisons between different 
parts of a network and 
provide a basic language for 
the description of network structure.

\begin{figure}[t!]
\centering
\epsfig{file=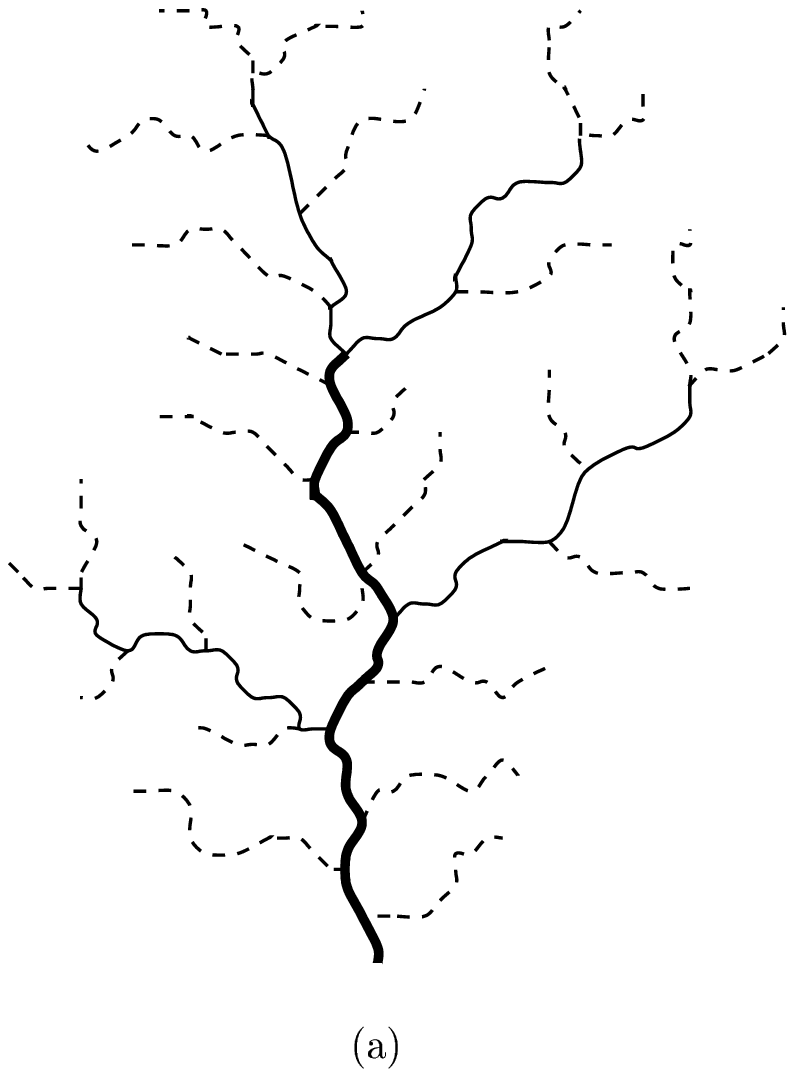,width=.3435\textwidth}
\qquad
\epsfig{file=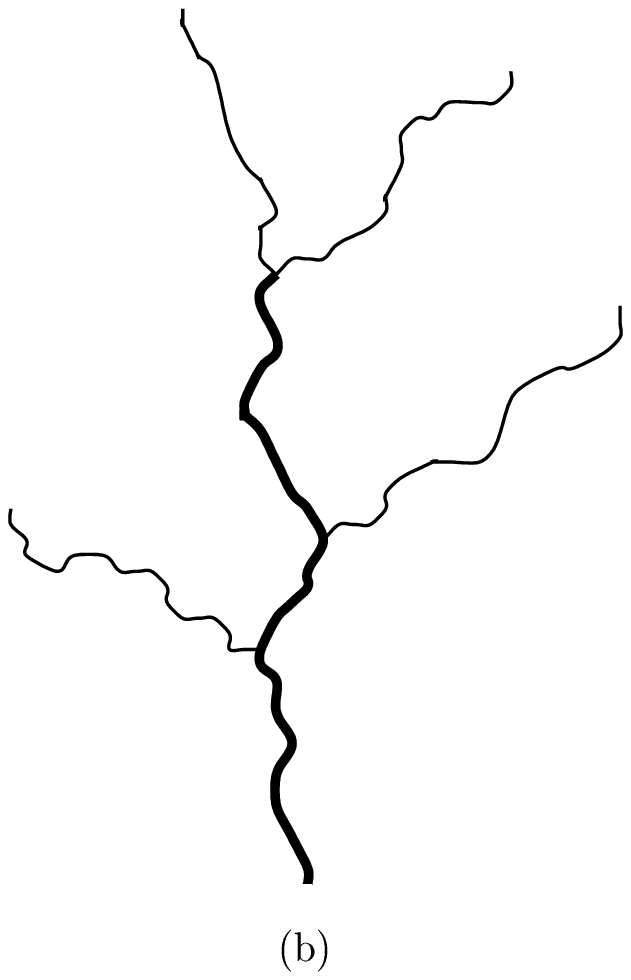,width=.27\textwidth}
\qquad\qquad
\epsfig{file=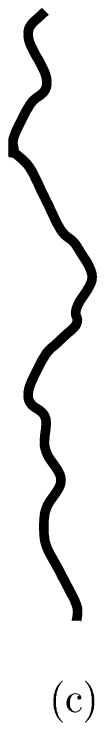,width=.0435\textwidth}
\caption{Horton-Strahler stream ordering.  (a) shows
the basic network.  (b) is created by removing all source streams 
from the network in (a), these same streams being denoted as
first order `stream segments'.  The new source streams in the pruned network of
(b) are labelled as second order stream segments and are themselves
removed to give (c), a third order stream segment.}
\label{fig:scaling.order}
\end{figure}

Here, we build our theory using
the most common ordering scheme,
one that was first introduced by
\Ai{Horton} in his seminal work on erosion~\cite{horton45}.
\Ai{Strahler} later improved this method~\cite{strahler57} 
and the resulting technique
is commonly referred to as Horton-Strahler stream 
ordering~\cite{rodriguez-iturbe97}.
The most natural description of this stream ordering, due to
\Ai{Melton}~\cite{melton59}, 
is based on an iterative pruning of a
tree representing a network as shown in Figure~\ref{fig:scaling.order}.
All source (or external) streams are pared away from the tree,
these being defined as the network's first order `stream segments'.  
A new tree is thus created along with a new collection of source streams
and these are precisely the second order
stream segments of the original network.  
The pruning and order identification
continues in like fashion
until only the trunk stream segment of the river network is left.
The overall order of the basin itself is identified
with the highest stream order present.

The usual and equivalent description details how
stream orders change at junctions~\cite{rodriguez-iturbe97}.  
When a stream segment of order $\om_1$
merges with a stream segment of order $\om_2$, the outgoing stream
will have an order of $\om$ given by
\begin{equation}
\om = \max(\om_1,\om_2) + \delta_{\om_1,\om_2}
\label{eq:scaling.streamordering}
\end{equation}
where $\delta$ is the Kronecker delta.
In other words, stream order
only increases when two stream segments of the same order come together
and, otherwise, the highest order is maintained by the outflowing
stream.

\section{Planform network quantities and scaling laws}
\label{sec:scaling.defs}
The results of this paper pertain
to networks as viewed in planform.
As such, any effects involving
relief, the vertical dimension, are ignored.  Nevertheless,
we show that a coherent theory of planform quantities may
still be obtained. This section defines the relevant quantities
and their various permutations
along with scaling laws observed to hold between them.
The descriptions of these laws will be short and
more detail will be provided in later sections.

The two essential features in river networks are basins
and the streams that drain them.  The two basic planform
quantities associated with these are drainage area
and stream length.  An understanding of the distribution
of these quantities is of fundamental importance in geomorphology. 
Drainage area, for example, serves as a measure of average discharge
of a basin
while its relationship with the length of the main 
stream gives a sense of how basins are shaped.

\subsection{General network quantities}
\label{sec:scaling.genquants}
\begin{figure}[tb!]
\centering
\epsfig{file=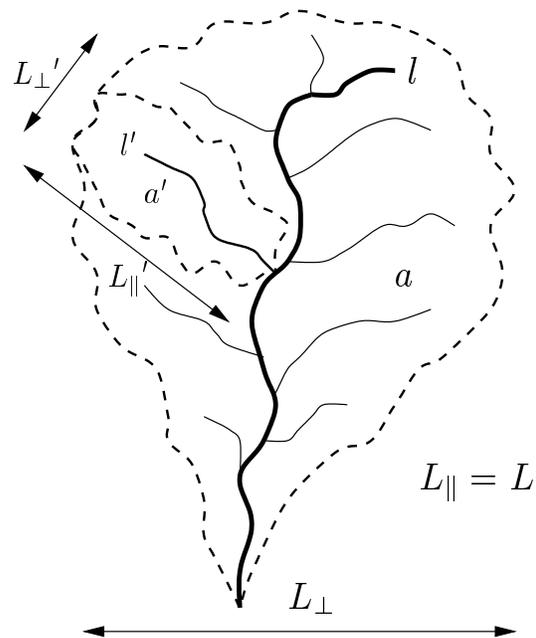}
\newline
\caption{A planform view of an example basin.  The main defining
parameters of a basin are $a$, the drainage area, 
$l$, the length of the main stream,
and $L_\parallel$ and $L_\perp$, the overall Euclidean dimensions.
The sub-basin with primed quantities demonstrates
that a basin exists at every point in a network.}
\label{fig:scaling.planformq}
\end{figure}

Figure~\ref{fig:scaling.planformq} shows a typical drainage basin.
The basin features are $a$, the area, $l$, the length of the
main stream, and $L_\parallel$ and $L_\perp$, the overall 
dimensions.  The main (or trunk) stream is the dominant
stream of the network---it is traced out by moving all the way upstream from 
the outlet to the start of a source stream by
choosing at each junction (or fork) the incoming stream with
the largest drainage area.  This is not to be confused with stream
segment length which only makes sense in the context of stream ordering.
We will usually write $L$ for $L_\parallel$.
Note that any point on a network has its own basin and
associated main stream.  The sub-basin
in Figure~\ref{fig:scaling.planformq} illustrates this and
has its own primed versions of $a$, $l$, $L_\parallel$ and $L_\perp$.
The scaling laws usually involve comparisons between basins of 
varying size.  These basins must be from the same landscape
and may or may not be contained within each other.

Several scaling laws connect these quantities.  
One of the most well known is Hack's law~\cite{hack57}.
Hack's law states that $l$
scales with $a$ as
\begin{equation}
l \sim a^{h}
\label{eq:scaling.hack}
\end{equation}
where $h$ is often referred to as Hack's exponent.
The important feature of Hack's law is that $h \neq 1/2$.   
In particular, it has been observed that for a reasonable
span of basin sizes that $0.57 < h < 0.60$~\cite{hack57,maritan96a,gray61,rigon96}. 
The actual range of this scaling is an unresolved issue 
with some studies demonstrating that very large basins 
exhibit the more expected scaling of 
$h=1/2$~\cite{mueller72,mosley73,mueller73}.
We simply show later that while the assumptions of this paper hold
so too does Hack's law.

Further comparisons of drainage basins of different sizes
yield scaling in terms of $L (= L_\parallel)$, the overall basin length.
Area, main stream length, and basin width are all 
observed to scale with $L$~\cite{maritan96a,tarboton88,labarbera89,tarboton90,labarbera90},
\begin{equation}
a \sim L^D, \qquad l \sim L^{{d}}, \qquad L_\perp \sim L^{H}.
\label{eq:scaling.scalingwithL}
\end{equation}

Turning our attention to the entire landscape,
it is also observed that histograms of stream lengths and basin areas reveal
power law distributions~\cite{maritan96a,rodriguez-iturbe97}:
\begin{equation}
P(a) \sim a^{-\tau}  \qquad  \mbox{and} \qquad P(l) \sim l^{-\gamma}.
\label{eq:scaling.powerlawdist}
\end{equation}

There are any number of other definable quantities and we
will limit ourselves to a few that are closely related to each other.  
We write $\lambda$ for the average distance from
a point on the network to the
outlet of a basin (along streams) and
$\Lambda$ for the unnormalized total of these distances.
A minor variation of these are $\tilde{\lambda}$ and $\tilde{\Lambda}$,
where only distances from junctions in the network to the outlet are included
in the averages.

The scaling law involving these particular 
quantities is Langbein's law~\cite{langbein47}
which states that
\begin{equation}
\Lambda \sim a^\beta.
\label{eq:scaling.langbeinslaw}
\end{equation}
Similarly, we have $\lambda \sim L^\varphi$, 
$\tilde{\Lambda} \sim a^{\tilde{\beta}}$
and $\tilde{\lambda} \sim L^{\tilde{\varphi}}$, \cite{maritan96a}.

\subsection{Network quantities associated with stream ordering}
With the introduction of stream ordering, 
a whole new collection of network quantities appear.
Here, we present the most important ones and discuss
them in the context of what
we identify as the principal structural laws
of river networks: Tokunaga's law and Horton's laws.

\subsubsection{Tokunaga's law}
Tokunaga's law concerns the set of ratios, $\{T_{\om,\om'}\}$,
first introduced by \Ai{Tokunaga}~\cite{tokunaga66,tokunaga78,tokunaga84,peckham95,newman97}.
These `Tokunaga ratios' represent the average 
number of streams of order $\om'$ flowing into a stream
of order $\om$ as side tributaries.
In the case of what we will call a `structurally self-similar network',
we have that $T_{\om,\om'} = T_{\om-\om'} = T_\nu$ where
$\nu = \om -\om'$ since quantities
involving comparisons between features at different scales should only
depend on the relative separation of those scales.
These $T_\nu$, in turn, are observed to be dependent 
such that~\cite{tokunaga66},
\begin{equation}
T_{\nu+1}/T_{\nu} = R_T
\label{eq:scaling.tokdef}
\end{equation}
where $R_T$ is a fixed constant for a given network.  
Thus, all of Tokunaga's ratios may
be specified by two fundamental parameters $T_1$ and $R_T$:
\begin{equation}
T_\nu = T_1 (R_T)^{\nu-1}.
\label{eq:scaling.toklaw}
\end{equation}
We refer to this last identity as Tokunaga's law.  

The network parameter $T_1$ is the average number
of major side tributaries per stream segment.  So for a
collection of stream segments of order $\om$, there will
be on average $T_1$ side tributaries of order $\om-1$
for each stream segment.  The second network parameter 
$R_T$ describes how numbers of
side tributaries of successively lower orders increase, again, on average.
As an example, consider that the network in 
Figure~\ref{fig:scaling.order} is
part of a much larger network for which $T_1 = 2$
and $R_T = 4$.
Figure~\ref{fig:scaling.order}~(b) shows that 
the third order stream segment has two major side tributaries of second order
which fits exactly with $T_1 = 2$
(Note that the two second order stream segments that come together to
create the third order stream segment are not side tributaries).
Figure~\ref{fig:scaling.order}~(a) further shows nine first 
order tributaries, slightly above
the average eight suggested by $T_2 = T_1 R_T^{1} = 8$.
Finally, again referring to Figure~\ref{fig:scaling.order}~(a),
there are $9/4 = 2.25$ first order tributaries for
each second order stream segment, not far from
the expected number $T_1 = 2$.

\subsubsection{Horton's laws}
\Ai{Horton} introduced several important measurements
for networks in conjunction with his stream ordering~\cite{horton45}.  
The first is the bifurcation ratio, $R_n$.  
This is the ratio of
the number $n_\omega$ of streams of order $\omega$
to the number $n_{\omega+1}$ of streams of order
$\omega + 1$ and is, moreover, observed
to be independent of $\omega$ over a large range.
There is next the stream length ratio,
$R_\ell = \bar{\ell}_{\om+1}/\bar{\ell}_{\om}$,
where $\bar{\ell}_{\om}$
is the average length of stream segments of order $\om$.
These lengths only exist within
the context of stream ordering.
In contrast to these are the main stream lengths,
which we have 
denoted by $l$ and described in section~\ref{sec:scaling.genquants}.
Main stream lengths are defined regardless of stream ordering 
and, as such,
are a more natural quantity.  Note that stream ordering gives
rise to a discrete set of basins, one for each junction in the network.
We therefore also have a set of basin areas and main stream lengths
defined at each junction.  Taking averages over basins of the
same order we have $\bar{a}_\om$ and $\bar{l}_\om$ to
add to the previously defined $\bar{\ell}_\om$ and $n_\om$.

The connection between the two measures of stream
length is an important,
if simple, exercise~\cite{scheidegger68c}.  Assuming 
$\bar{\ell}_{\om+1} = R_{\ell} \bar{\ell}_{\om}$ holds for
all $\omega$, one has
\begin{equation}
\bar{l}_{\om} = \sum_{i=1}^{\om} \bar{\ell}_{i}
= \sum_{i=1}^{\om} (R_{\ell})^{i-1}\bar{\ell}_{1}
= \bar{l}_{1}\frac{(R_{\ell})^{\om}-1}{R_{\ell}-1}
\label{eq:scaling.lconnect}
\end{equation}
where $\bar{l}_1 = \bar{\ell}_1$ has been used.  
Since typically $R_\ell > 2$~\cite{kirchner93}, 
$\bar{l}_{\om+1}/\bar{l}_\om \rightarrow R_{\ell}$ rapidly.
For $\om=4$ and $R_\ell = 2$, the error is only three per cent.
On the other hand, starting with the assumption that
main stream lengths satisfy Horton's law of stream lengths for all $\omega$
implies that the same is true for stream segments.

Thus, for most calculations,
Horton's law of stream lengths may involve either stream
segments or main streams and, for convenience, we will assume
that the law is fully satisfied by the former.
Furthermore, this small calculation suggests
that studies involving only third- or fourth-order networks
cannot be presumed to have reached
asymptotic regimes of scaling laws.  
We will return to this point throughout the paper.

Schumm~\cite{schumm56a} is attributed with the concrete introduction
of a third and final law that was also suggested by Horton.
This last ratio is for drainage areas
and states that $R_a = \bar{a}_{\om+1}/\bar{a}_{\om}$.
We will later show in section~\ref{sec:scaling.Ra=Rn} that
our assumptions lead to the result that $R_a \equiv R_n$.
At this stage, however, we write Horton's
laws as the three statements
\begin{equation}
\frac{n_{\om}}{n_{\om+1}} = R_n, 
\
\frac{\bar{\ell}_{\om+1}}{\bar{\ell}_{\om}} = R_{\ell}, 
\ \mbox{and} \
\frac{\bar{a}_{\om+1}}{\bar{a}_{\om}} = R_a.
\label{eq:scaling.hortslaws}
\end{equation}

A summary of all of the scaling laws presented in this
section is provided in Table~\ref{tab:scaling.scalinglaws}.
Empirically observed values for the relevant exponents
and ratios are presented in Table~\ref{tab:scaling.values}.
\begin{table}[t]
\begin{center}
\begin{tabular}{cl} 
\tbf{Law:} & \tbf{Name or description:}\\ \hline
$T_{\nu} = T_1 (R_T)^{\nu-1}$ & \tbf{Tokunaga's law}\\
$l \sim L^{{d}}$ & \tbf{self-affinity of single channels}\\
$n_{\om+1}/n_{\om} = R_n$ & Horton's law of stream numbers\\
$\bar{\ell}_{\om+1}/\bar{\ell}_{\om} = R_{\ell}$ & Horton's law of stream segment lengths\\
$\bar{l}_{\om+1}/\bar{l}_{\om} = R_{\ell}$ & Horton's law of main stream lengths\\
$\bar{a}_{\om+1}/\bar{a}_{\om} = R_a$ & Horton's law of stream areas\\
$l \sim a^h$ & Hack's law\\
$a \sim L^D$ & scaling of basin areas\\
$L_\perp \sim L^H$ & scaling of basin widths\\
$P(a) \sim a^{-\tau}$ & probability of basin areas\\
$P(l)\sim l^{-\gamma}$ & probability of stream lengths\\
$\Lambda \sim a^\beta$ & Langbein's law\\
$\lambda \sim L^\varphi$ & variation of Langbein's law\\
$\tilde{\Lambda} \sim a^{\tilde{\beta}}$ & as above\\
$\tilde{\lambda} \sim L^{\tilde{\varphi}}$ & as above\\
\end{tabular}
\caption{A general list of scaling laws for river networks.  
All laws and quantities are defined in section~\ref{sec:scaling.defs}.
The principal finding of this paper
is that these scaling laws
follow from the first two relations, Tokunaga's law
(structural self-similarity) and
the self-affinity of single channels, and the assumption
of uniform drainage density
(defined in section~\ref{sec:scaling.assumptionsdd}).}
\label{tab:scaling.scalinglaws}
\end{center}
\end{table}

\subsection{Scheidegger's random networks}
To end this introductory section, we detail some of
the features of
the random network model of Scheidegger~\cite{scheidegger67b,scheidegger90}.
Although originally defined without reference to a real surface,
Scheidegger networks may be obtained from a completely uncorrelated landscape
as follows.  Assign a random height between 0 and 1 at every point
on a triangular lattice and then tilt the lattice so that no
local minima (lakes) remain.  Scheidegger networks are then traced
out by following paths of steepest descent.

Surprisingly, these networks still exhibit all of the scaling
laws observed in real networks.  It thus provides an important
point in `network space' and
accordingly, also provides an elementary test for any theory
of scaling laws.
Exact analytical results for various exponents are known due to
the work of \Ai{Takayasu} \etal\ on the aggregation of particles
with injection~\cite{takayasu88,takayasu89a,takayasu89b,takayasu90,takayasu91,huber91}.  
While there are no analytic results for the Tokunaga ratio
$T_1$ or the Horton ratios $R_n$ and $R_{\ell}$, our own simulations
show that these stream order laws are strictly obeyed.
Table~\ref{tab:scaling.values} lists the
relevant exponents and their values for the \sche\ model along
with those found in real networks.

\begin{table}[t]
\begin{center}
\begin{tabular}{ccc}
\tbf{Quantity:} & \tbf{Scheidegger:}    & \tbf{Real networks:}  \\ \hline
$R_n$           & $5.20 \pm .05$        & 3.0--5.0 \cite{abrahams84}\\
$R_a$           & $5.20 \pm .05$        & 3.0--6.0 \cite{abrahams84}\\
$R_\ell$        & $3.00 \pm .05$        & 1.5--3.0 \cite{abrahams84}\\
$T_1$           & $1.30 \pm .05$        & 1.0--1.5 \cite{tokunaga78}\\
${d}$           & $1$                   & $1.1 \pm 0.01$ \cite{maritan96a}\\
$D$             & $3/2$                 & $1.8 \pm 0.1$ \cite{maritan96a}\\
$h$             & $2/3$                 & 0.57--0.60 \cite{maritan96a}\\
$\tau$          & $4/3$                 & $1.43 \pm 0.02$ \cite{maritan96a}\\
$\gamma$        & $3/2$                 & $1.8 \pm 0.1$ \cite{rigon96}\\
$\varphi$       & $1$                   & $1.05 \pm 0.01$ \cite{maritan96a}\\
$H$             & $1/2$                 & 0.75--0.80 \cite{maritan96a}\\ 
$\beta$         & $5/3$                 & 1.56 \cite{langbein47}\\
$\varphi$       & $1$                   & $1.05 \pm 0.01$ \cite{maritan96a}\\
\end{tabular}
\caption{Ratios and scaling exponents for Scheidegger's random
network model
and real networks.  For Scheidegger's model,
exact values are known due to the work of \Ai{Takayasu} \etal\
\protect 
\cite{takayasu88,takayasu89a,takayasu89b,takayasu90,takayasu91,huber91}
and approximate results are taken from our own simulations.
For real networks, the references given are generally the most
recent and further appropriate references may be found within them
and also in section~\ref{sec:scaling.defs}.}
\label{tab:scaling.values}
\end{center}
\end{table}

\section{Assumptions}
\label{sec:scaling.assumptions}
We start from three basic assumptions about the structure
of river networks: structural self-similarity, self-affinity
of individual streams and uniformity of drainage density.
We define these assumptions and their relevant parameters
and then discuss their mutual consistency.  We end with a discussion
of the correspondence between the laws of  Tokunaga and Horton.
It should be stressed that while we make a case for each
assumption there is also considerable 
proof to ponder in the pudding that these
ingredients create.

\subsection{Structural self-similarity}
\label{sec:scaling.assumptionsss}
Our first assumption is that networks are structurally self-similar.
It has been observed that river networks exhibit self-similarity
over a large range of scales~\cite{mandelbrot83,tarboton88,rodriguez-iturbe97}.
Naturally, the physical range of this self-similarity is restricted to
lie between two scales.  The large scale cutoff is the overall size of the
landscape and the small scale cutoff is of the order of
the characteristic separation of channels~\cite{montgomery92}.

In order to quantify this phenomenon,
we look to laws of network structure such as Tokunaga's law
and Horton's laws of stream number and length.
We demonstrate in the following section
that these descriptions are mutually consistent within the
context of our third assumption, uniformity of drainage density.  
Thus, we may assume a network where both Tokunaga's and Horton's
laws hold.  For convenience, we write these laws
as if they hold for all orders down to the first order.  Any actual
deviations from these laws for low orders will not affect the
results since we are interested in how laws behave for increasing
stream order.

\subsection{Self-affinity of individual streams}
\label{sec:scaling.assumptionsdl}
Our second assumption is that individual streams are
self-affine curves possessing a dimension ${d}>1$,
as introduced in equation~\req{eq:scaling.scalingwithL}.
Empirical support for this premise 
is to be found in~\cite{tarboton88,labarbera89,tarboton90,maritan96a,rodriguez-iturbe97,tarboton89}.
In reality, this is at best a weak fractality with measurements
generally finding ${d}$ to be 
around $1.1$~\cite{maritan96a}.
We assume ${d}$ to be constant throughout a given network,
true for each stream independent of order.

In general, it is most reasonable to
consider this in the sense of a growing fractal: stream
length $l$ will grow like $L^{{d}}$ where $L$ is the overall length
of a box containing a portion of a stream.  
So, rather than examine one fixed section
of a stream, we take larger and larger pieces of it.
Moreover, this is the most reasonable method for actually 
measuring ${d}$ for a real network.  

\subsection{Uniform drainage density}
\label{sec:scaling.assumptionsdd}
Our third and final assumption is that drainage density
is uniform throughout a network.  For a given basin,
the drainage density, $\rho$, is a measure of the average area drained per
unit length of stream by overland flow (i.e., excluding
contributions from tributary streams).
Its usual form is that given by Horton~\cite{horton45}:
\begin{equation}
\rho = \frac{\sum\ell}{a}
\label{eq:scaling.ddensity}
\end{equation}
where, for a given basin, $\sum\ell$ represents
the summed total length of all stream segments of all orders and $a$ is
the drainage area.  More generally, one can in the same way
measure a local drainage
density for any connected sections of a network within a landscape.
Such sections should cover a region at least $\ell_1$ in diameter, the
typical length of a first order stream.
Drainage density being uniform means that the variation of
this local drainage density is negligible.
There is good support in the literature for the uniformity
of drainage density in real 
networks~\cite{hack57,shreve67,haggett69,gardiner73,morisawa62,devries94}
while there are some suggestions that it may vary slightly with
order~\cite{hack57,tokunaga78}.

Uniform drainage density may also be interpreted 
as the observation that the average distance between channels is
roughly constant throughout a landscape~\cite{horton45,rodriguez-iturbe97},
an estimate of this distance being simply $1/\rho$.
This is due to the fact that
there is a finite limit to the channelization of a landscape
determined by a combination of soil properties, climate and so on.
Implicit in this assumption is that the channel network has
reached its maximum extension into a landscape~\cite{shreve67,glock31}.
Indeed, In the bold words of \Ai{Glock}~\cite{glock31}, 
we are considering river networks at
the ``time of completed territorial conquest.''  Furthermore,
\Ai{Shreve}~\cite{shreve67} notes that drainage density would be uniform in
a ``mature topography developed in a homogeneous environment.''

Importantly, our third assumption connects the planform 
description to the surface
within which the network lies.
Computationally, the uniformity of drainage density allows for the use of the
length of a stream as a proxy for drainage area~\cite{devries94}.
Further, the average distance between streams
being roughly constant implies that, on average,
tributaries are spaced evenly along a stream.

\section{Tokunaga's law and Horton's laws are equivalent}
This section demonstrates an equivalence between
Tokunaga's law and Horton's two laws of stream number and 
stream length in the case of a landscape with uniform
drainage density.

\subsection{From Tokunaga's law to Horton's laws}
\Ai{Tokunaga} has shown that
Horton's law for stream numbers follows from 
Tokunaga's law 
(given in equation~\req{eq:scaling.toklaw})~\cite{tokunaga78,peckham95}.
This follows from the observation that $n_\om$, the number
of streams of order $\om$, in a basin of order $\Om$ may be expressed as
\begin{equation}
n_\om = 2n_{\om+1} + \sum_{\nu=1}^{\Om-\om} T_\nu n_{\om+\nu}.
\label{eq:scaling.nwtok}
\end{equation}
The $2n_{\om+1}$ accounts for the fact that each order $\om+1$
stream is initiated by the confluence of two streams of
order $\om$.  
Presuming Tokunaga's law,
a simple analysis of equation~\req{eq:scaling.nwtok} 
shows that in the limit of large $\Om$, the ratio $n_\om/n_{\om+1}$
does indeed approach a constant.
This leads to an expression for the Horton ratio $R_n$ in
terms of the two Tokunaga parameters $T_1$ and $R_T$
(first obtained by \Ai{Tokunaga} in~\cite{tokunaga78}):
\begin{equation}
2 R_n = (2 + R_T + T_1) + \left[(2 + R_T + T_1)^2 - 8R_T\right]^{1/2}.
\label{eq:scaling.tokhortlink1}
\end{equation}

\Ai{Tokunaga}'s work has been recently generalized by \Ai{Peckham} 
who deduces links to the other 
Horton ratios $R_{\ell}$ and $R_a$~\cite{peckham95}.
In contrast to the purely algebraic calculation of $R_n$,
these results require the step of equating topological properties 
to metric basin quantities.  In determining $R_{\ell}$,
Peckham uses the number of side tributaries to a stream
as an estimate of stream segment length. 
This is based on the assumption that tributaries are evenly
spaced.
As discussed in section~\ref{sec:scaling.assumptionsdd},
this even spacing of tributaries follows for networks
with uniform drainage density.
Therefore, we may write, after \Ai{Peckham}, that
\begin{equation}
\bar{\ell}_\om \propto 1+\sum_{\nu=1}^{\om-1}{T_\nu}
\label{eq:scaling.tokhortlink2a}
\end{equation}
where the dimension of length absent on the right-hand side
is carried by an appropriate constant of proportionality.
This sum is simply the total number of tributaries that, on average, enter
a stream of order $\om$.  The number of lengths of stream between
tributaries is then simply one more in number.

Using Tokunaga's law (equation~\req{eq:scaling.toklaw})
we find that
\begin{equation}
\bar{\ell}_{\om+1}/\bar{\ell}_{\om}
 = R_T \left( 1 + O (R_T)^{-\omega} \right),
\label{eq:scaling.tokhortlink2b}
\end{equation}
obtaining \Ai{Horton}'s stream length ratio with the simple
identification:
\begin{equation}
R_{\ell} = R_T
\label{eq:scaling.tokhortlink2}
\end{equation}
and we will use $R_{\ell}$ in place of $R_T$ throughout the rest of
the paper.
As already noted
we will see that $R_a \equiv R_n$ for landscapes where drainage density
is uniform.
This redundancy means that there are only
two independent Horton ratios, $R_{\ell}$ and $R_n$, which sits well
with the two independent quantities required for Tokunaga's law,
$T_1$ and $R_T$.  
Presupposing this result, we can invert 
equations~\req{eq:scaling.tokhortlink1} and~\req{eq:scaling.tokhortlink2}
to obtain Tokunaga's parameters from the two independent Horton ratios:
\begin{eqnarray}
R_T & = & R_{\ell} \\
T_1 & = & R_n - R_{\ell} - 2 + 2R_{\ell}/R_n.
\label{eq:scaling.tokhortinv}
\end{eqnarray}

\subsection{From Horton's laws to Tokunaga's law}
We now provide an heuristic argument to show that  Tokunaga's
law in the form of equation~\req{eq:scaling.toklaw} follows
from Horton's laws of stream number and length
and uniform drainage density.  Note that even though
we have shown in equations~\req{eq:scaling.tokhortlink1},
\req{eq:scaling.tokhortlink2},
and~\req{eq:scaling.tokhortinv} that the parameters of
Tokunaga's law and those of Horton's laws may be obtained
from each other, it is not a priori clear that this result
would be true.  Indeed, Tokunaga's law contains more direct
information about network structure than Horton's laws and
it is the additional constraint of uniform drainage density
that provides the key.

\begin{figure}[tb!]
\centering
\epsfig{file=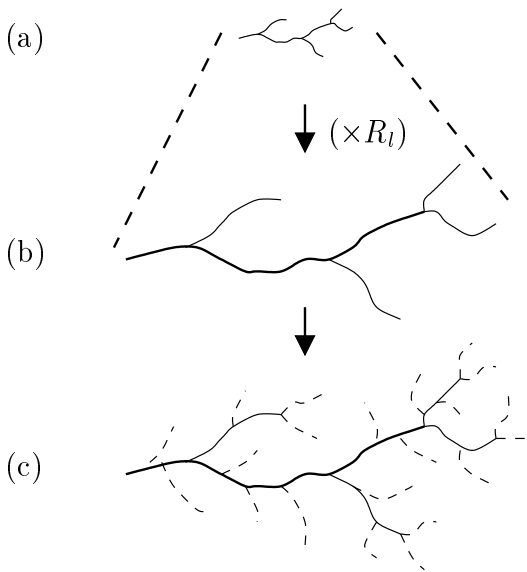}
\caption{An example rescaling of a basin to demonstrate how
Tokunaga's law follows from Horton's laws and uniform drainage
density.  In the first step from (a) to (b), the streams of the small network are rescaled
in length by a factor of $R_{\ell}$.  The second step from (b) to
(c) demonstrates that
for drainage density to remain constant and uniform, a sufficient
number of first order tributaries must be added.}
\label{fig:scaling.hort2tok}
\end{figure}

Consider a stream of order $\om$ along with its
side tributaries of order $\om' = 1$ through $\om' =\om-1$, 
the numbers of which are given by the usual $T_{\nu}$ where
$\nu = \om-\om'$
(see Figure~\ref{fig:scaling.hort2tok}).  Since 
the presumed adherence to Horton's laws
implies that a network is self-similar we need only consider
the form of the $T_\nu$ and not the more general $T_{\om',\om}$.
Now, again since networks are self-similar, a typical
stream of order $\om + 1$ can be obtained by scaling up the picture
of this order
$\om$ stream.  As per Horton's law of stream lengths, this is done by
increasing the length of each stream by a factor of $R_{\ell}$
(Figure~\ref{fig:scaling.hort2tok} (a) 
becomes Figure~\ref{fig:scaling.hort2tok} (b)).

However, since order $\om'$ streams become $\om'+1$ streams
in this rescaling, the picture in Figure~\ref{fig:scaling.hort2tok} (b)
is missing first order streams.  Also, the average distance
between tributaries has grown by a factor of $R_{\ell}$.  Therefore,
to retain the same drainage density, 
an extra $(R_{\ell}-1)$ first order streams
must be added for each link (one more than the number of
tributaries) along this new order $\om + 1$ stream (Figure~\ref{fig:scaling.hort2tok} (c)).
Since the number of first order streams is now given
by $T_{\om+1}$ we have
\begin{equation}
T_{\om+1} = (R_{\ell}-1)\left(\sum_{\nu=1}^{\om} T_\nu + 1 \right).
\label{eq:scaling.hort2tok1}
\end{equation}
It may be simply checked that this equation is satisfied,
for large $\om$, by Tokunaga
ratios given by equation~\req{eq:scaling.toklaw}.
Thus, Horton's laws of stream number and stream length and
the uniform drainage density are seen to imply Tokunaga's law.

In general, Horton's ratios rather than
the parameters of Tokunaga's law will 
be the most useful parameters in what follows.
In particular, we will see that the
two independent quantities $R_n$ and $R_{\ell}$ will be needed
only in the form $\ln R_n / \ln R_{\ell}$.  All other
exponents will be expressible as algebraic combinations of
$\ln R_n / \ln R_{\ell}$ and ${d}$, the fractal dimension of an individual
stream.

Furthermore, example (or modal) values for the parameters of Horton and
Tokunaga are~\cite{tokunaga78,kirchner93}
\begin{equation}
T_1 = 1, \ R_T = R_\ell = 2, \ \mbox{and} \ R_n = 4.
\label{eq:scaling.typical}
\end{equation}
The parameters have been chosen so as to satisfy
the inversion relations of equation~\req{eq:scaling.tokhortinv}.
As shown in Table~\ref{tab:scaling.values},
real networks provide some variation around these modal values.
These will be used as rough checks of accuracy throughout
the rest of the paper.

\section{Hack's law}
One of the most intriguing scalings found in river networks
is Hack's law~\cite{hack57} which relates main stream
length to basin area as $l \sim a^h$.
This equation has been empirically shown to hold true for a large range of
drainage basin sizes on many field sites~\cite{rodriguez-iturbe97}.
The salient
feature is that for smaller basins~\cite{rigon96},
 $h$ is typically found to be in the range $(0.56,0.60)$,
whereas $0.5$ would be expected from 
simple dimensional analysis~\cite{rodriguez-iturbe97}.

It should be emphasized that Hack's law is only true
on average as are, for that matter, Tokunaga's law and Horton's laws.
An extension of Hack's law to
a more natural statistical description of the connection between
stream lengths and drainage areas was suggested by \Ai{Maritan} 
\etal~\cite{maritan96a} with some further developments to be found
in~\cite{dodds98d}.

\subsection{Horton's other law of stream numbers}
In order to obtain Hack's law, we will use the uniformity
of drainage density to estimate the area of an order $\Om$ basin
by calculating the total length of streams within the same basin.
So we simply need the typical length and number
of each stream order present.
Taking the length of a source stream, $\bar{\ell}_1$, to be the
finest resolution of the network and the basic unit of length, 
the length of a stream segment
of order $\om$ is $\bar{\ell}_{\om}=(R_{\ell})^{\om-1}\bar{\ell}_1$.
However, in finding the frequency of such streams we find that some care
must be taken for the following reasons.

Horton's law of stream numbers is potentially misleading in
that it suggests, at first glance, 
that within a basin of order $\om$ there should be
one stream of order $\om$,
$R_n$ streams of order $\om-1$, $R_n^2$ streams of order $\om-2$
and so on.  Indeed, many calculations involving Horton's laws
use this assumption~\cite{labarbera89,rodriguez-iturbe97,devries94,rosso91}.

But Horton's $R_n$ actually provides the ratio of
the number of streams of consecutive orders as totalled for 
a \textit{whole basin}.
To illustrate this fact, consider streams of order $\om$ and $\om+1$
within a basin of order $\Om \gg \om$.  As Tokunaga's law makes
clear, streams of order $\om$ are not all found within sub-basins
of order $\om+1$.  Indeed, a certain number of
 order $\om$ streams will be tributaries
to streams of order greater than $\om+1$ (see the example network
of Figure~\ref{fig:scaling.order} (a)).  Tokunaga's law shows that
we should in fact expect $T_1+2$ rather than than $R_n$
streams of order $\om$ entering into a stream of order $\om+1$.
For the typical values $T_1=1$ and $R_n=4$ in~\req{eq:scaling.typical}
this is a substantial error.

We proceed then to find a corrected version of Horton's law of stream numbers.
Returning to equation~\req{eq:scaling.nwtok}, we see that it
is only valid in the limit $\Om \rightarrow \infty$.  Defining
$n'(\om,\Om)$ as the actual number of streams of order $\om$ \textit{within}
a basin of order $\Om$, we have
\begin{equation}
n'(\om,\Om) = 2n'(\om+1,\Om) + \sum_{\nu=1}^{\Om-\om} T_\nu n'(\om+\nu,\Om).
\label{eq:scaling.nwtokbetter}
\end{equation}
This equation may be exactly solved.  Considering
the above expression for $n'(\om,\Om)$ and the corresponding one
for $n'(\om+1,\Om)$ we can reduce this to a simple difference equation,
\begin{equation}
n'(\om,\Om) = (2+R_{\ell}+T_1)n'(\om+1,\Om) -2R_{\ell} n'(\om+2,\Om)
\label{eq:scaling.nwtokdiffeq}
\end{equation}
which has
solutions of the form $\mu^k$.
Applying the constraints that $n'(\Om,\Om)=1$
and $n'(\Om-1,\Om)=T_1+2$, we obtain
\begin{equation}
n'(\om,\Om) = c (\mu_{+})^{\Om-\om} + (1-c) (\mu_{-})^{\Om-\om}
\label{eq:scaling.nwtokdiffeqsoln}
\end{equation}
where
\begin{equation}
2\mu_{\pm} = (2+R_{\ell}+T_1) \pm \left[(2 + R_\ell + T_1)^2 - 8R_\ell \right]^{1/2}
\label{eq:scaling.mudef}
\end{equation}
and 
\begin{equation}
c=R_n(R_n-R_{\ell})/(R_n^2-2R_{\ell}).
\label{eq:scaling.cdef}
\end{equation}
Note that $R_n = \mu_+$ and
we will use the notation $R_n^\ast$ in place of $\mu_-$.
This observation regarding Horton's law of stream numbers was 
first made by \toku~\cite{tokunaga66} and later
by \Ai{Smart}~\cite{smart67}.
In particular, Tokunaga noted that this would 
explain the deviation of Horton's law for the highest orders 
of a basin, a strong motivation for his work.

We can now define an effective Horton ratio, ${R_n}'(\om,\Om)$ as follows:
\begin{eqnarray}
{R_n}'(\om,\Om) & = & n'(\om-1,\Om)/n'(\om,\Om) \nonumber\\
 & = & R_n \left( 1 + O(R_n^\ast/R_n)^{(\Om-\om)} \right)
\end{eqnarray}
The typical values of Horton's ratios in~\req{eq:scaling.typical}
give $R_n^\ast = 1$. 
In this case, ${R_n}'(\om,\Om)$ converges rapidly to $R_n$ with an
error of around one per cent for $\om=\Om-3$.

\subsection{Hack's law}
\label{sec:scaling.hackslaw}
As discussed in section~\ref{sec:scaling.assumptionsdd}, 
an estimate of total drainage area of a basin is
given by the total length of all streams within the basin.
Summing over all stream orders and using the numbers $n'(\om,\Om)$
given by equations~\req{eq:scaling.nwtokdiffeqsoln}
and~\req{eq:scaling.mudef} we have that
\begin{eqnarray}
\bar{a}_\Om & \propto & 
\sum_{\om=1}^{\Om} 
n'(\om,\Om) (R_{\ell})^{\om-1} \nonumber \\
 & = & c_1(R_n)^{\Om} + c_2(R_{\ell})^{\Om} - c_3(R^\ast_n)^{\Om}
\label{eq:scaling.areawRBRL}
\end{eqnarray}
where $c_1=c/(R_n-R_\ell)$, $c_3=(1-c)/(R_\ell-R^\ast_n)$ and $c_2=c_3-c_1$
with $c$ being given in equation~\req{eq:scaling.cdef}.
Slightly more complicated is the estimate of $\bar{a}(\om,\Om)$,
the drainage area of a basin of order $\om$ within a
basin of order $\Om$:
\begin{eqnarray}
\bar{a}(\om,\Om)
& \propto & 1/n'(\om,\Om) \sum_{\om'=1}^{\om} 
n'(\om',\Om) (R_{\ell})^{\om'-1}  \nonumber \\
 & = & 1/n'(\om,\Om)
\left[
c_1 (R_n)^\Om(1-(R_\ell/R_n)^\om) \right. \nonumber \\
 & & +
\left. c_3 (R_\ell)^\om (R_n^\ast)^{\Om-\om}(1-(R_n^\ast/R_\ell)^\om)
\right].
\label{eq:scaling.areawRBRLsub}
\end{eqnarray}
Now, for $1 \ll \om \ll \Om$ (typically, $3<\om<\Om-2$ is
sufficient), this expression is well approximated as
\begin{equation}
\bar{a}(\om,\Om) \sim (R_n)^\om.
\label{eq:scaling.areaapprox}
\end{equation}
since $R_n > R_{\ell} > R_n^\ast$.

Thus, we have also shown here that $R_a \equiv R_n$.
While it is true that we would have obtained the same with
a naive use of Horton's laws, we have both made the derivation
thorough and established the correction terms found in
equation~\req{eq:scaling.areawRBRLsub}.  
This will be investigated further in the next section.

Finally, using this result
and the estimate $\bar{l}_{\om} \propto (R_{\ell})^{\om}$ from 
equation~\req{eq:scaling.lconnect}, it follows that
\begin{equation} 
\bar{l}_{\om} \propto (R_{\ell})^{\om} = (R_n)^{\om \ln R_{\ell}/\ln R_n}
\sim (\bar{a}_{\om})^{\ln R_{\ell}/\ln R_n}
\label{eq:scaling.RBRLhack}
\end{equation}
which is precisely Hack's law.
Comparing equations~\req{eq:scaling.RBRLhack}
and~\req{eq:scaling.hack}, Hack's exponent is found in
terms of the Horton ratios $R_n$ and $R_{\ell}$ as
\begin{equation} 
h = \frac{\ln R_{\ell}}{\ln R_n}.
\label{eq:scaling.hackRBRL}
\end{equation}
There is one minor caveat to the derivation in \req{eq:scaling.RBRLhack}
and, for that matter, to most other derivations in this paper.
Equation~\req{eq:scaling.RBRLhack} only 
holds for the characteristic areas and lengths
$\bar{a}_\om$ and $\bar{l}_\om$.  
Since these quantities grow exponentially
with $\om$, the derivation gives evenly spaced points on
a log-log plot lying on a straight line.  
Clearly, this would
indicate that the actual relationship is continuous
and linear on a log-log plot. 
Indeed, there is no obvious reason that a network
would prefer certain lengths and areas.  
The averaging of stream lengths and areas brought about by the
imposition of stream ordering
necessarily removes all information contained in higher order statistics.
Motivated by this observation,
generalizations of the laws of Tokunaga, Horton and Hack to
laws of distributions rather than averages is in progress~\cite{dodds98d}.

\section{There are only two Horton ratios}
\label{sec:scaling.Ra=Rn}
In deriving Hack's law in the previous section we
obtained from equation~\req{eq:scaling.areaapprox}
that $R_a \equiv R_n$.  This redundancy in Horton's laws 
is implicit in, amongst others, the works of Horton~\cite{horton45}
and Hack~\cite{hack57} but has never been stated outright.
As noted previously, \Ai{Peckham} also obtains a similar result for
a topological quantity, the number of source streams in a basin,
that is used as an estimate of area.
Thus, we see that for a landscape with uniform
drainage density, Horton's laws are fully specified by only
two parameters $R_n$ and $R_\ell$.  This further supports
our claim that Tokunaga's law and Horton's laws are equivalent
since we have shown that
there is an invertible transformation between
$(T_1,R_T)$, the parameters of Tokunaga's law, and $(R_n,R_\ell)$
(equations~\req{eq:scaling.tokhortlink1}, \req{eq:scaling.tokhortlink2}
and~\req{eq:scaling.tokhortinv}).  
In this section, we present data from real networks that support
the finding $R_n = R_a$.  We also address reported cases that do not
conform to this result and consider a possible explanation in
light of the correction terms
established in equation~\req{eq:scaling.areawRBRL}.

\begin{figure}[tb!]
\centering
\epsfig{file=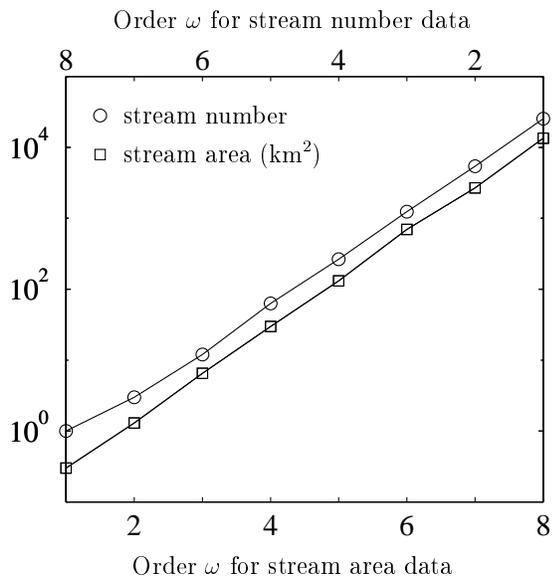,width=.4\textwidth}
\caption{Average area and stream number as functions of stream order
for Kentucky River, Kentucky (data taken from Peckham 
\protect \cite{peckham95}).  The stream number data is reversed
for simpler comparison with the area data.  The Horton ratios are
estimated to be $R_n \approx R_a = 4.65 \pm 0.05$.}
\label{fig:scaling.kentucky}
\end{figure}

\begin{figure}[htb!]
\centering
\epsfig{file=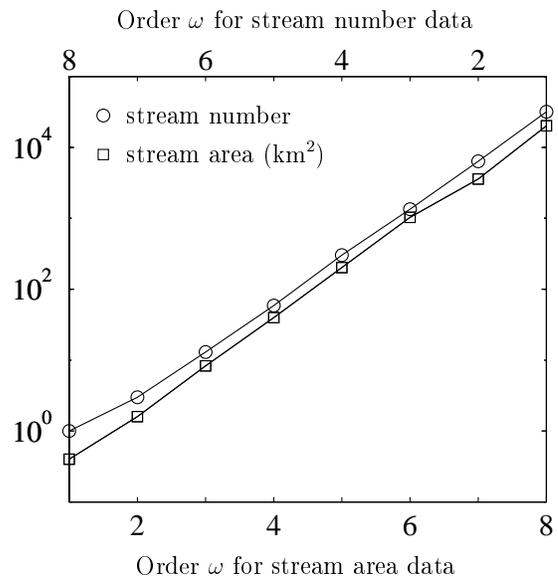,width=.4\textwidth}
\caption{Average area and stream number as functions of stream order
for Powder River, Wyoming (data taken from Peckham \protect \cite{peckham95}).
Here the ratios are $R_n \approx R_a = 4.55 \pm 0.05$.}
\label{fig:scaling.powder}
\end{figure}

Excellent agreement for the result $R_n = R_a$
in real networks is to be found in the data of \Ai{Peckham}~\cite{peckham95}.
The data is taken from an analysis of digital elevation models (DEM's) for
the Kentucky River, Kentucky and the Powder River, Wyoming.
Figure~\ref{fig:scaling.kentucky} shows average area and stream
number plotted as a function of order for the Kentucky River while
Figure~\ref{fig:scaling.powder} shows the same for the
Powder river.  Note that stream number has been plotted against
decreasing stream order to make the comparison clear.
The exponents $R_a$ and $R_n$ are
indistinguishable in both cases.  For the Kentucky river, 
$R_n \approx R_a = 4.65 \pm 0.05$ and for the Powder river,
$R_n \approx R_a = 4.55 \pm 0.05$.
Also of note here is that the same equality is well satisfied by \sche's model
where numerical simulations yield values of $R_a = 5.20 \pm 0.05$ and 
$R_n = 5.20 \pm 0.05$.

Note the slight deviation from a linear
form for stream numbers for large $\om$ in both cases.
This upwards concavity is as predicted by the
modified version of Horton's law of stream numbers for
a single basin, equation~\req{eq:scaling.nwtokdiffeqsoln}.

At the other extreme, the fit
for both stream areas and stream numbers extends to $\om = 1$.
While this may seem remarkable,
it is conceivable that at the resolution of the DEM's used, some
orders of smaller streams may have been removed by coarse-graining.  Thus,
$\om=1$ may actually be, for example, a third order stream.  Note that
such a translation in the value of $\om$ does not affect the
determination of the ratios as it merely results in the change
of an unimportant multiplicative constant.
If $\om_r$ is the true order and
$\om = \om_r - m$, where $m$ is some integer, then, for example,
\begin{equation}
n_\om \propto (R_n)^{\om} \sim (R_n)^{\om_r - m} 
= \mbox{const} \times (R_n)^{\om_r}.
\label{eq:scaling.omok}
\end{equation}
This is only a rough argument as coarse-graining does not necessarily
remove all streams of low orders.

At odds with the result that $R_n \equiv R_a$ are past
measurements that uniformly find $R_a > R_n$ at a number of sites.
For example, Rosso \etal\ in~\cite{rosso91} examine eight river
networks and find
$R_a$ to be on average 40 \% greater than $R_n$.
Clearly, this may be solely due to one or more of the our assumptions
not being satisfied.  The most likely would be that drainage density
is not uniform.   However, the limited size of the data sets points
to a stronger possibility which we now discuss.

In the case of~\cite{rosso91}, the networks considered are 
all third or fourth order basins 
with one exception of a fifth order basin.
As shown by equation~\req{eq:scaling.areawRBRL}, if
Horton's laws of stream number and length are exactly followed
for all orders, Horton's law of area is not obeyed for lower orders.
Moreover, the former are most likely asymptotic relations themselves.
It is thus unsatisfactory to make estimates of 
Horton's ratios from only three or four data points taken from
the lowest order basins.  Note that
the Kentucky and Powder rivers are both eighth order networks
and thus provide a sufficient range of data.

\begin{figure}[tb!]
\centering
  \epsfig{file=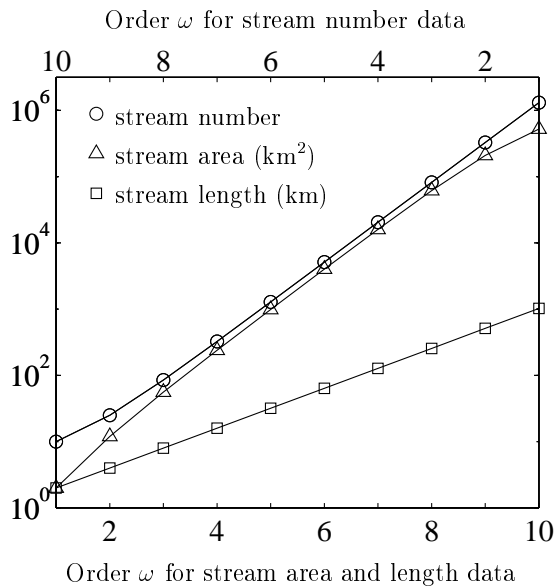,width=.4\textwidth}
\caption{An explanation for the empirical finding that $R_n < R_a$.
Fitting a line to the stream area for only low $\om$ would result
in an overestimate of its asymptotic slope.  For stream number,
its slope would be underestimated.}
\label{fig:scaling.RAfit}
\end{figure}

We consider more precisely how the corrections to the scaling of
area given in equation~\req{eq:scaling.areawRBRLsub} would
affect the measurement of the Horton ratios.
Figure~\ref{fig:scaling.RAfit} shows an example of
how stream number, length and area might vary with $\om$.  It is
assumed, for the sake of argument, 
that stream number and length scale exactly as per Horton's
laws and that area behaves as in equation~\req{eq:scaling.areawRBRLsub}, 
satisfying Horton's law of area
only for higher values of $\om$.  The plot
is made for the example values $R_n=4$ and $R_{\ell}=2$.
The prefactors are chosen
arbitrarily so the ordinate is of no real significance.

A measurement  of $R_a$ from
a few data points in the low $\om$ range will
overestimate its asymptotic value as will a similar measurement
of $R_n$ underestimate its true value.
Estimates of $R_n$ and $R_a$ from a simple least squares fit for various ranges
of data are provided in Table~\ref{tab:scaling.RaRnmess}.

\begin{table}
\begin{tabular}{r|cccc}
$\om$ range& $1,2,3$ & $1,2,3,4$ & $1,2,3,4,5$ & $4,5,6,7,8$ \\ 
\hline
$R_n$ & 2.92 & 3.21 & 3.41 & 3.99 \\
$R_a$ & 5.29 & 4.90 & 4.67 & 4.00 \\
\end{tabular}
\caption{Values of Horton ratios obtained
from least squares estimates of slopes for data represented
in \protect Figure~\ref{fig:scaling.RAfit}.
The range indicates
the data points used in the estimate of the slopes.  The
ratios obtained from
the low order data demonstrate substantial error whereas those
obtained from the middle data essentially give the true values
of $R_n = R_a = 4$.}
\label{tab:scaling.RaRnmess}
\end{table}

Thus, the validity of the methods and 
results from past work are cast in some doubt.
A reexamination of data which
has yielded $R_a \gg R_n$ appears warranted with
an added focus on drainage density.  Moreover, it is 
clear that networks of a much higher order must be studied to
produce any reasonable results.

\section{Fractal dimensions of networks: a revision}
A number of papers and works over the past decade have analyzed the
relationships that exist between Horton's laws and two
fractal dimensions used to describe river 
networks~\cite{tarboton88,labarbera89,tarboton90,rosso91,feder88,labarbera94,stark97}.
These are $D$, the dimension which describes the scaling of the total mass
of a network, and ${d}$, the fractal dimension of individual
streams that comprises one of our assumptions.  In this section,
we briefly review these results and point out several inconsistencies.
We then provide a revision that fits within the context of
our assumptions.

Our starting point is
the work of \Ai{La Barbera} and \Ai{Rosso}~\cite{labarbera89}
which was improved by \Ai{Tarboton} \etal\ to give~\cite{tarboton90}
\begin{equation}
D = {d}\frac{\ln R_n}{\ln R_{\ell}}.
\label{eq:scaling.DdRBRL}
\end{equation}
We find this relation to be correct but that the assumptions and derivations
involved need to be redressed.
To see this, note that equation~\req{eq:scaling.DdRBRL}
was shown to follow from two observations.  The
first was the estimation of $N(\bar{\ell}_1)$, the number of
boxes of size $\bar{\ell_1} \times \bar{\ell_1}$ required
to cover the network~\cite{labarbera89}: 
\begin{equation}
N(\bar{\ell}_1) \sim (\bar{\ell}_1)^{-\ln R_n/\ln R_{\ell}}
\label{eq:scaling.Nl1}
\end{equation}
where $\bar{\ell_1}$ is 
the mean length of first order stream segments.
Note that Horton's laws were directly
used in this derivation rather than the correctly modified law of stream
numbers for single basins (equation~\req{eq:scaling.nwtokdiffeqsoln}).  
Nevertheless, the results are the same asymptotically.
The next was the inclusion of our second assumption, that single channels are 
self-affine~\cite{tarboton90}.  Thus, it was claimed,
$\bar{\ell}_1 \sim \delta^{-{d}}$ where
$\delta$ is now the length of the measuring stick.  Substitution
of this into equation~\req{eq:scaling.Nl1} gave
\begin{equation}
N(\delta) \sim \delta^{-{d} \ln R_n/\ln R_{\ell}},
\label{eq:scaling.Ndelta}
\end{equation}  
yielding the stated expression for $D$, equation~\req{eq:scaling.DdRBRL}.

\begin{figure}[tb!]
\centering
\epsfig{file=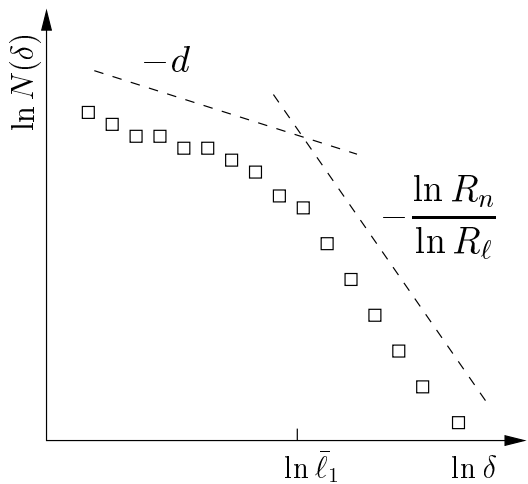}
\caption{A schematic representing the problems associated with
measuring the fractal dimension of \textit{a single river network}.
Here, the box counting method is assumed and
$\delta$, which has the units of length, is the side length of 
the $N(\delta)$ boxes needed to cover the network.
For box sizes much greater than $\bar{\ell}_1\times \bar{\ell}_1$, only the network
structure is detected while for box sizes smaller
than $\bar{\ell}_1\times \bar{\ell}_1$, the measurement picks out 
the fractal dimension of individual streams.
Some deviation towards the scaling suggested by
equation~\req{eq:scaling.Ndelta} may occur between these two limits.}
\label{fig:scaling.crossover}
\end{figure}

However, there is one major assumption in this work that needs to be more
carefully examined.  The network is assumed to be of infinite order, i.e.,
one can keep finding smaller and smaller streams.  As we have
stated, there is a finite limit to the extension of any real network. 
The possible practical effects of this are pictorially represented in 
Figure~\ref{fig:scaling.crossover}.  Consider that the network in question
is of actual order $\Om$.  Then there are three possible scaling regimes.
Firstly, for a ruler of length $\delta \gg \bar{\ell}_1$,
only the network structure may be detected, given that individual
streams are almost one-dimensional.  Here, the scaling exponent will be
$\ln R_n/\ln R_{\ell}$.  Next, as $\delta$ decreases,
the fractal structure of individual streams may come into play and 
the exponent would approach that of equation~\req{eq:scaling.Ndelta}.
Depending on the given network, this middle section may not even be
present or, if so, perhaps only as a small deviation as depicted.
Finally, the contribution due to the overall network structure must vanish
by the time $\delta$ falls below $\bar{\ell}_1$.  From this point on, the
measurement can only detect the fractal nature of individual streams
and so the exponent must fall back to ${d}$.

We therefore must rework this derivation of equation~\req{eq:scaling.DdRBRL}.
As suggested in the definition of ${d}$ in
section~\ref{sec:scaling.assumptionsdl}, it is more reasonable to
treat networks as growing fractals.  Indeed, since there is a finite
limit to the extent of channelization of a landscape, there is
a lower cut-off length scale beyond which most network quantities have
no meaning.  The only reasonable way to examine scaling behavior
is to consider how these quantities change with increasing basin size.
This in turn can only be done
by comparing different basins of increasing order as opposed
to examining one particular basin alone.  

With this in mind, the claim that equation~\req{eq:scaling.DdRBRL}
is the correct scaling can be argued as follows.  Within
some basin of order $\Om$, take a sub-basin of order $\om$. 
Consider $N(\om)$, the number of 
boxes of side length $\bar{\ell}_1$ required to cover the sub-network.
This is essentially given by the total length of all the streams
in the network.  This is given by
the approximation of equation~\req{eq:scaling.areaapprox} and so we have
that $N(\om) \propto (R_n)^\om$.
Using the fact that $\bar{\ell}_\om = 
(R_{\ell})^{\om-1}\bar{\ell}_1$ we then have
that $N(\om) \propto (\ell_\om/\bar{\ell}_1)^{\ln R_n/ \ln R_{\ell}}$.
The difference here is that $\bar{\ell}_1$ is fixed
and pertains to the actual first
order streams of the network.  By assumption, we have that 
$\ell_\om \propto L^{{d}}$ and thus
\begin{equation}
N(L) \propto L^{{d} \ln R_n/\ln R_{\ell}},
\label{eq:scaling.Dcorrect}
\end{equation}
which gives the same value for $D$ as equation~\req{eq:scaling.DdRBRL}.

There are two other relations involving fractal dimensions
that also need to be reexamined.
Firstly \Ai{Rosso} \etal~\cite{rosso91} found that
\begin{equation}
{d} = 2\frac{\ln R_{\ell}}{\ln R_a}.
\label{eq:scaling.dRLRA}
\end{equation}
Combining equations~\req{eq:scaling.DdRBRL}
and~\req{eq:scaling.dRLRA}, they then obtained
\begin{equation}
D = 2\frac{\ln R_n}{\ln R_a}.
\label{eq:scaling.dRBRA}
\end{equation}
However, equation~\req{eq:scaling.dRLRA} and
hence equation~\req{eq:scaling.dRBRA} are both incorrect.

There is a simple explanation for this discrepancy.
In deriving equation~\req{eq:scaling.dRLRA},
\Ai{Rosso} \etal\ make the assumption that $h={d}/2$,
a hypothesis first suggested by Mandelbrot~\cite{mandelbrot83}.
In arriving at the relation $h={d}/2$,
Mandelbrot states in~\cite{mandelbrot83} that
``$(\mbox{basin area})^{1/2}$ should be proportional to
(distance from source to mouth as the crow flies).''
In other words, $a \propto L^{1/2}$.  However, as noted in
equation~\req{eq:scaling.scalingwithL}, observations
of real networks
show that $a \propto L^D$ where $D < 2$~\cite{maritan96a}.
Furthermore, on examining the result $h = \ln R_\ell / \ln R_n$
with the expression
for $D$ in equation~\req{eq:scaling.DdRBRL} we see that
\begin{equation}
h = \frac{{d}}{D},
\label{eq:scaling.hackdD}
\end{equation}
which suggests that this hypothesis is valid only when $D=2$.
Consider also the test case of the \sche\ model where $h = 2/3$,
$D = 3/2$ and ${d} = 1$ (see Table~\ref{tab:scaling.values}).  
Using these values, we see that equation ~\req{eq:scaling.hackdD}
is exactly satisfied while the relation $h={d}/2$
gives $h=1/2 \neq 2/3$.

Now, if $h={d}/D$ is used in place of
$h={d}/2$ in deriving equation~\req{eq:scaling.dRLRA} then
equation~\req{eq:scaling.DdRBRL} is recovered.  
It also follows that equation~\req{eq:scaling.dRBRA}
simplifies to the statement $R_a = R_n$, further demonstrating
the consistency of our derivations.
Thus, the two equations~\req{eq:scaling.dRLRA} and~\req{eq:scaling.dRBRA}
become redundant and
the only connection between Horton's ratios and network
dimensions is given by equation~\req{eq:scaling.DdRBRL}.

An important point is that $D < 2$
\textit{does not imply} that drainage basins are not
space filling.  This exponent shows how basin area changes
when comparing different basins with different values of $L$,
i.e., $a \propto L^D$.  
Any given single basin
has of course a fractal dimension of 2.  The equating of
the way basin sizes change with the actual dimension of
any one particular basin is a confusion evident in the 
literature (see, for example, \cite{tarboton88}).
Incorporating the effects of measuring basin area
with boxes of side length $\delta$ in the relation $a \propto L^D$
would lead to the form
\begin{equation}
a_L(\delta) \propto \delta^{-2} L^{D},
\label{eq:scaling.adeltaLH}
\end{equation}
where the subscript $L$ has been used to emphasize that
different values of $L$ correspond to different basins.
Thus, for any given basin (i.e., for fixed $L$), the area scales with
$\delta$ while for a fixed $\delta$, areas of different basins
scale as per equation~\req{eq:scaling.scalingwithL}.

It should also be emphasized that the relationship found here 
between Hack's exponent and the fractal dimensions
${d}$ and $D$ is one that is explicitly derived from the
assumptions made.
The observation that basin areas scale non-trivially
with $L$ follows from
these starting points and thus there is no need to assume it here.

\section{Other scaling laws}
We now address three remaining sets of scaling laws.
These are probability distributions for areas and stream
lengths, scaling of basin shape and Langbein's law.

As introduced in equation~\req{eq:scaling.powerlawdist},
probability distributions for $a$ and $l$ are observed
to be power law with exponents $\tau$ and $\gamma$~\cite{rodriguez-iturbe97}.
Both of these laws have previously been derived from Horton's laws.
\Ai{De Vries} \etal\ \cite{devries94} found a relationship between
$\tau$, $R_n$ and $R_{\ell}$ but did not include ${d}$ in their
calculations while \Ai{Tarboton} \etal~\cite{tarboton88} 
obtained a result for $\gamma$ that did
incorporate ${d}$.  

Again, both of these derivations use Horton's laws directly rather than
the modified version of equation~\req{eq:scaling.nwtokdiffeqsoln}.
Asymptotically, the same results are obtained from both approaches,
\begin{equation}
\tau = 2 - \frac{\ln R_{\ell}}{\ln R_n}
\quad \mbox{and} \quad
\gamma = \frac{\ln R_n}{\ln R_{\ell}}.
\label{eq:scaling.tau}
\end{equation}
Using the form of the Hack exponent found in 
equation~\req{eq:scaling.hackdD} and equation~\req{eq:scaling.DdRBRL},
further connections between these exponents are found:
\begin{equation}
\tau = 2 - h
\quad \mbox{and} \quad \gamma = \frac{1}{h}.
\label{eq:scaling.tauh}
\end{equation}

One important outcome concerns
the fact that only one of the exponents
of the triplet $(h,\tau,\gamma)$ is independent.
Previously, for the particular case of directed networks, 
this has been shown by \Ai{Meakin} \etal~\cite{meakin91}
and further developed by \Ai{Colaiori} \etal~\cite{colaiori97}.
Directed networks are those networks in which all flow
has a non-zero positive component in a given direction.
In a different setting,
\Ai{Cieplak} \etal\ also arrive at this same conclusion
for what they deem to be the separate cases of self-similar
and self-affine networks although their assumptions are
that $d < 1$ and $D < 2$ are mutually exclusive contrary 
to empirical observations~\cite{cieplak98a}.
In the case of non-directed networks,
\Ai{Maritan} \etal\ have found one scaling relation for these three exponents,
$\gamma = 1 + (\tau - 1)/h$ and, therefore, that two
of these three exponents are independent.  They further
noted that $\tau = 2 - h$ is an ``intriguing result''
suggested by real data~\cite{maritan96a}.
In the present context,
we have obtained this reduction of description
in a very general way
with, in particular, no assumption regarding the directedness of the networks.

The scaling of basin shapes has been addressed already but
it remains to show how it simply follows from our assumptions
and how the relevant exponents are related.
It is enough to show that this scaling follows from Hack's law.
Now, the area of a basin is related
to the longitudinal length $L$ and the width $L_\perp$
by $a = L_\perp L$, while the main stream length scales
by assumption like $l \sim L^{{d}}$.  Hence,
\begin{eqnarray}
l \sim a^h & \Rightarrow & L^{{d}} \sim (L_\perp L)^h \nonumber \\
           & \Rightarrow & L_\perp \sim L^{{d}/h - 1} = L^{D - 1}
\label{eq:scaling.self-aff}
\end{eqnarray}
where the fact that $h = {d}/D$ has been used.  
Comparing this to equation~\req{eq:scaling.scalingwithL} 
we obtain the scaling relation
\begin{equation}
H = D - 1.
\label{eq:scaling.HD}
\end{equation}

The last set of exponents
we discuss are those relating to
Langbein's law~\cite{langbein47}.
Langbein found that $\tilde{\Lambda}$, the sum of the
distances (along streams) from stream junctions to the
outlet of a basin, scales with the area
of the basin.  Recently, Maritan \etal\ \cite{maritan96a} 
introduced the quantity $\lambda$, which is an average
of Langbein's $\tilde{\Lambda}$ except now the sum is taken over
all points of the network.  Citing the
case of self-organized critical networks, they
made the claim that
\begin{equation}
\lambda \propto L^\varphi.
\label{eq:scaling.elldef}
\end{equation}
Further, they assumed that $\varphi = {d}$ although it was
noted that there is no clear reason why this may be so since
there are evident differences in definition ($\lambda$ involves
distances downstream while ${d}$ involves distances
upstream).  We find this scaling relation to hold in
the present framework.
We further consider the two related quantities $\Lambda$ and
$\tilde{\lambda}$, respectively the sum over all
points and the average over all junctions of distances
along streams to the basin outlet.

The calculations are straightforward and follow the
manner of previous sections.  We first calculate
$\lambda(\om,\Om)$, the typical distance to the outlet from
a stream of order $\om$ in an order $\Om$ basin.  Langbein's
$\tilde{\Lambda}$, for example, is then obtained as 
$\sum_{\om=1}^\Om n(\om,\Om) \lambda(\om,\Om)$.
We find the same scaling behavior regardless of
whether sums are taken over all points or all junctions.
Specifically we find
\begin{equation}
\Lambda \sim \tilde{\Lambda} \sim a^{1+\ln R_\ell/\ln R_n}
\quad \mbox{and} \quad
\lambda \sim \tilde{\lambda} \sim L^d
\label{eq:scaling.lambascaling}
\end{equation}
yielding the scaling relations
\begin{equation}
\beta = \tilde{\beta} = {1+\ln R_\ell/\ln R_n}
\quad \mbox{and} \quad
\tilde{\varphi} = \varphi = {d}.
\label{eq:scaling.varphid}
\end{equation}
Note that the second pair of scaling relations admit
other methods of measuring ${d}$.  The large amount of
averaging inherent in the definition of the quantity $\lambda$
would suggest that it is a more robust method for measuring
${d}$ than one based on measurements of
the sole main stream of the basins.

\Ai{Maritan} \etal~\cite{maritan96a} 
provide a list of real world measurements for various exponents
upon which several comments should be made.  Of particular
note is the relationship between $\tau = 2 - h$.  This
is well met by the cited values $1.41 < \tau < 1.45$
and $0.57 < h < 0.60$.  Also reasonable is the estimate
of $h$ given by ${d}/D$ ($D=\phi$ in their notation)
which is $0.58<h<0.65$.  

The values of $\gamma$ and $\varphi$, however,
do not work quite so well.  The latter does not match ${d}$
within error bars, although they are close in absolute value
with $\varphi = 1.05\pm0.01$
and ${d}=1.10\pm0.01$.  The length distribution exponent $\gamma$
may be found via 3 separate routes: $\gamma = 1/h = D/{d} = 1/(2-\tau)$.
The second and third equalities have been noted to be well satisfied
and so any one of the 3 estimates of $\gamma$ may be used.
Take, for example, the range $0.58 < h < 0.59$, which falls within
that given by $h=2-\tau$, $h = {d}/D$ and the range given for $h$ itself.
This points to the possibility that the measured range
$1.8 < \gamma < 1.9$
is too high, since using $\gamma = 1/h$ yields $\gamma = 1.74 \pm .02$.
Also of note is that Maritan \etal's own scaling relation
$\gamma = 1 + (\tau-1)/h$ would suggest $\gamma = 1.74 \pm .05$.

Better general agreement with the scaling relations
is to be found in \cite{rigon96} in which
\Ai{Rigon} \etal\ detail specific values of $h$, $\tau$
and $\gamma$ for some thirteen river networks.  Here, the 
relations $\tau = 2-h$
and $\gamma = 1/h$ are both well satisfied.  Comparisons
for this set of data show that, on average and given the
cited values of $h$, 
both $\tau$ and $\gamma$ are overestimated by only 2 per cent.

\section{Concluding remarks}
We have demonstrated that the various laws, exponents and parameters
found in the description of river networks follow from a few
simple assumptions.  Further, all quantities are expressible
in terms of two fundamental numbers.  These are a ratio of
logarithms of Horton's ratios, $\ln R_n/\ln R_{\ell}$, and 
the fractal dimension of individual streams, ${d}$.  There
are \textit{only two} independent parameters in network scaling laws.
These Horton ratios were shown to be equivalent to Tokunaga's law
in informational content with the attendant assumption of
uniform drainage density.  Further support for this observation
is that both the Horton and Tokunaga 
descriptions depend on two parameters each and an invertible
transformation between them exists
(see equations~\req{eq:scaling.tokhortlink1}, 
\req{eq:scaling.tokhortlink2} and~\req{eq:scaling.tokhortinv}).  
A summary of the connections found between the various exponents
is presented in Table~\ref{tab:scaling.laws}.  

It should be emphasized that the importance of laws like
that of Tokunaga and Horton in the description of networks
is that they provide explicit structural information.  
Other measurements such as the power law probability distributions
for length and area provide little information about
how a network fits together.
Indeed, information is lost
in the derivations as the Horton ratios cannot
be recovered from knowledge of $\ln R_n/ \ln R_{\ell}$ and ${d}$ only.

The basic assumptions of this work
need to be critically examined.
Determining how often they hold and why they hold
will follow through to a greater understanding 
of all river network laws.  One vital part of any river network
theory that is lacking here is
the inclusion of the  effects of relief, the third dimension.
Another is the dynamics of network growth: why do mature river
networks exhibit a self-similarity that gives rise to these
scaling laws with these particular values of exponents?
Also, extensive studies of variations in drainage density
are required.
The assumption of its uniformity plays a critical role in
the derivations and needs to be reexamined.
Lastly, in those cases where these assumptions are valid, the
scaling relations gathered here provide a powerful method
of cross-checking measurements.

Finally, we note that work of a similar nature has recently
been applied to biological networks~\cite{west97}.  
The assumption analogous to network self-similarity used
in the biological setting is considerably weaker as it requires
only that the network is a hierarchy.  A principle of 
minimal work is then claimed to constrain this hierarchy to be
self-similar.  It is conceivable that a similar approach may
be found in river networks.  However, a generalization of the concept
of a hierarchy and perhaps stream ordering 
needs to be developed since a `Tokunagic network'
is not itself a simple hierarchy.

\begin{table}
\begin{center}
\begin{tabular}{cl} 
\tbf{law:} & \tbf{parameter in terms of $R_n$, $R_{\ell}$ and ${d}$:} \\ \hline
$T_{\nu} = T_1 (R_T)^{\nu-1}$ & $T_1 = R_n -R_\ell - 2 + 2R_\ell/R_n$\\
         & $R_T = R_\ell$ \\
$l \sim L^{{d}}$ & ---\\
$n_{\om+1}/n_{\om} = R_n$ & ---\\
$\bar{\ell}_{\om+1}/\bar{\ell}_{\om} = R_{\ell}$ & ---\\
$\bar{l}_{\om+1}/\bar{l}_{\om} = R_{\ell}$ & ---\\
$\bar{a}_{\om+1}/\bar{a}_{\om} \sim R_a$ & $R_a = R_n$ \\
$l \sim a^h$ & $h = \ln R_{\ell}/ \ln R_n$ \\
$a \sim L^D$ & $D = {d} \ln R_n/\ln R_{\ell}$ \\
$L_\perp \sim L^H$ & $H = {d} \ln R_n/\ln R_{\ell} - 1$ \\
$P(a) \sim a^{-\tau}$ & $\tau = 2 - \ln R_{\ell}/ \ln R_n$ \\
$P(l)\sim l^{-\gamma}$ & $\gamma = \ln R_n/ \ln R_{\ell}$ \\
$\Lambda \sim a^\beta$ & $\beta = 1 + \ln R_{\ell}/\ln R_n$ \\
$\lambda \sim L^\varphi$ & $\varphi = {d}$ \\
$\tilde{\Lambda} \sim a^{\tilde{\beta}}$ & $\tilde{\beta} = 1 + \ln R_{\ell}/\ln R_n$ \\
$\tilde{\lambda} \sim L^{\tilde{\varphi}}$ & $\tilde{\varphi} = {d}$ \\
\end{tabular}
\caption{
  Summary of scaling laws and the scaling relations found
  between the various exponents.  
  Compare with table~\ref{tab:scaling.scalinglaws}. 
}
\label{tab:scaling.laws}
\end{center}
\end{table}

\section*{Acknowledgements}
We are grateful to R.\ Pastor-Satorras, J.\ Pelletier, G.\ West, 
J.\ Weitz and K.\ Whipple for useful discussions.  The work
was supported in part by NSF grant EAR-9706220.


\begin{thebibliography}{10}
\expandafter\ifx\csname bibnamefont\endcsname\relax
  \def\bibnamefont#1{#1}\fi
\expandafter\ifx\csname bibfnamefont\endcsname\relax
  \def\bibfnamefont#1{#1}\fi
\expandafter\ifx\csname url\endcsname\relax
  \def\url#1{\texttt{#1}}\fi
\expandafter\ifx\csname urlprefix\endcsname\relax\def\urlprefix{URL }\fi
\expandafter\ifx\csname bibinfo\endcsname\relax \def\bibinfo#1#2{#2}\fi
\expandafter\ifx\csname eprint\endcsname\relax \def\eprint#1{#1}\fi

\bibitem{mandelbrot83}
\bibinfo{author}{\bibfnamefont{B.~B.} \bibnamefont{Mandelbrot}},
  \emph{\bibinfo{title}{The Fractal Geometry of Nature}}
  (\bibinfo{publisher}{Freeman}, \bibinfo{address}{San Francisco},
  \bibinfo{year}{1983}).

\bibitem{horton45}
\bibinfo{author}{\bibfnamefont{R.~E.} \bibnamefont{Horton}},
  \bibinfo{journal}{Bull. Geol. Soc. Am}
  \textbf{\bibinfo{volume}{56}}(\bibinfo{number}{3}), \bibinfo{pages}{275}
  (\bibinfo{year}{1945}).

\bibitem{langbein47}
\bibinfo{author}{\bibfnamefont{W.~B.} \bibnamefont{Langbein}},
  \bibinfo{journal}{U.S. Geol. Surv. Water-Supply Pap.}
  \textbf{\bibinfo{volume}{W 0968-C}}, \bibinfo{pages}{125}
  (\bibinfo{year}{1947}).

\bibitem{strahler52}
\bibinfo{author}{\bibfnamefont{A.~N.} \bibnamefont{Strahler}},
  \bibinfo{journal}{Bull. Geol. Soc. Am} \textbf{\bibinfo{volume}{63}},
  \bibinfo{pages}{1117} (\bibinfo{year}{1952}).

\bibitem{hack57}
\bibinfo{author}{\bibfnamefont{J.~T.} \bibnamefont{Hack}},
  \bibinfo{journal}{U.S. Geol. Surv. Prof. Pap.}
  \textbf{\bibinfo{volume}{294-B}}, \bibinfo{pages}{45} (\bibinfo{year}{1957}).

\bibitem{tarboton88}
\bibinfo{author}{\bibfnamefont{D.~G.} \bibnamefont{Tarboton}},
  \bibinfo{author}{\bibfnamefont{R.~L.} \bibnamefont{Bras}}, \bibnamefont{and}
  \bibinfo{author}{\bibfnamefont{I.}~\bibnamefont{Rodr\'{\i}guez-Iturbe}},
  \bibinfo{journal}{Water Resour. Res.}
  \textbf{\bibinfo{volume}{24}}(\bibinfo{number}{8}), \bibinfo{pages}{1317}
  (\bibinfo{year}{1988}).

\bibitem{labarbera89}
\bibinfo{author}{\bibfnamefont{P.}~\bibnamefont{La~Barbera}} \bibnamefont{and}
  \bibinfo{author}{\bibfnamefont{R.}~\bibnamefont{Rosso}},
  \bibinfo{journal}{Water Resour. Res.}
  \textbf{\bibinfo{volume}{25}}(\bibinfo{number}{4}), \bibinfo{pages}{735}
  (\bibinfo{year}{1989}).

\bibitem{tarboton90}
\bibinfo{author}{\bibfnamefont{D.~G.} \bibnamefont{Tarboton}},
  \bibinfo{author}{\bibfnamefont{R.~L.} \bibnamefont{Bras}}, \bibnamefont{and}
  \bibinfo{author}{\bibfnamefont{I.}~\bibnamefont{Rodr\'{\i}guez-Iturbe}},
  \bibinfo{journal}{Water Resour. Res.}
  \textbf{\bibinfo{volume}{26}}(\bibinfo{number}{9}), \bibinfo{pages}{2243}
  (\bibinfo{year}{1990}).

\bibitem{maritan96a}
\bibinfo{author}{\bibfnamefont{A.}~\bibnamefont{Maritan}},
  \bibinfo{author}{\bibfnamefont{A.}~\bibnamefont{Rinaldo}},
  \bibinfo{author}{\bibfnamefont{R.}~\bibnamefont{Rigon}},
  \bibinfo{author}{\bibfnamefont{A.}~\bibnamefont{Giacometti}},
  \bibnamefont{and}
  \bibinfo{author}{\bibfnamefont{I.}~\bibnamefont{Rodr\'{\i}guez-Iturbe}},
  \bibinfo{journal}{Phys. Rev. E}
  \textbf{\bibinfo{volume}{53}}(\bibinfo{number}{2}), \bibinfo{pages}{1510}
  (\bibinfo{year}{1996}).

\bibitem{leopold62}
\bibinfo{author}{\bibfnamefont{L.~B.} \bibnamefont{Leopold}} \bibnamefont{and}
  \bibinfo{author}{\bibfnamefont{W.~B.} \bibnamefont{Langbein}},
  \bibinfo{journal}{U.S. Geol. Surv. Prof. Pap.}
  \textbf{\bibinfo{volume}{500-A}}, \bibinfo{pages}{1} (\bibinfo{year}{1962}).

\bibitem{shreve66}
\bibinfo{author}{\bibfnamefont{R.~L.} \bibnamefont{Shreve}},
  \bibinfo{journal}{J. Geol.} \textbf{\bibinfo{volume}{74}},
  \bibinfo{pages}{17} (\bibinfo{year}{1966}).

\bibitem{howard71b}
\bibinfo{author}{\bibfnamefont{A.~D.} \bibnamefont{Howard}},
  \bibinfo{journal}{Geogr. Anal.} \textbf{\bibinfo{volume}{3}},
  \bibinfo{pages}{29} (\bibinfo{year}{1971}).

\bibitem{stark91}
\bibinfo{author}{\bibfnamefont{C.~P.} \bibnamefont{Stark}},
  \bibinfo{journal}{Nature} \textbf{\bibinfo{volume}{352}},
  \bibinfo{pages}{405} (\bibinfo{year}{1991}).

\bibitem{meakin91}
\bibinfo{author}{\bibfnamefont{P.}~\bibnamefont{Meakin}},
  \bibinfo{author}{\bibfnamefont{J.}~\bibnamefont{Feder}}, \bibnamefont{and}
  \bibinfo{author}{\bibfnamefont{T.}~\bibnamefont{J{\o}ssang}},
  \bibinfo{journal}{Physica A} \textbf{\bibinfo{volume}{176}},
  \bibinfo{pages}{409} (\bibinfo{year}{1991}).

\bibitem{willgoose91a}
\bibinfo{author}{\bibfnamefont{G.}~\bibnamefont{Willgoose}},
  \bibinfo{author}{\bibfnamefont{R.~L.} \bibnamefont{Bras}}, \bibnamefont{and}
  \bibinfo{author}{\bibfnamefont{I.}~\bibnamefont{Rodr\'{\i}guez-Iturbe}},
  \bibinfo{journal}{Water Resour. Res.}
  \textbf{\bibinfo{volume}{27}}(\bibinfo{number}{7}), \bibinfo{pages}{1685}
  (\bibinfo{year}{1991}).

\bibitem{willgoose91c}
\bibinfo{author}{\bibfnamefont{G.}~\bibnamefont{Willgoose}},
  \bibinfo{author}{\bibfnamefont{R.~L.} \bibnamefont{Bras}}, \bibnamefont{and}
  \bibinfo{author}{\bibfnamefont{I.}~\bibnamefont{Rodr\'{\i}guez-Iturbe}},
  \bibinfo{journal}{Earth Surf. Proc. Landforms} \textbf{\bibinfo{volume}{16}},
  \bibinfo{pages}{237} (\bibinfo{year}{1991}).

\bibitem{willgoose91}
\bibinfo{author}{\bibfnamefont{G.}~\bibnamefont{Willgoose}},
  \bibinfo{author}{\bibfnamefont{R.~L.} \bibnamefont{Bras}}, \bibnamefont{and}
  \bibinfo{author}{\bibfnamefont{I.}~\bibnamefont{Rodr\'{\i}guez-Iturbe}},
  \bibinfo{journal}{Water Resour. Res.}
  \textbf{\bibinfo{volume}{27}}(\bibinfo{number}{7}), \bibinfo{pages}{1671}
  (\bibinfo{year}{1991}).

\bibitem{kramer92}
\bibinfo{author}{\bibfnamefont{S.}~\bibnamefont{Kramer}} \bibnamefont{and}
  \bibinfo{author}{\bibfnamefont{M.}~\bibnamefont{Marder}},
  \bibinfo{journal}{Phys. Rev. Lett.}
  \textbf{\bibinfo{volume}{68}}(\bibinfo{number}{2}), \bibinfo{pages}{205}
  (\bibinfo{year}{1992}).

\bibitem{leheny93}
\bibinfo{author}{\bibfnamefont{R.~L.} \bibnamefont{Leheny}} \bibnamefont{and}
  \bibinfo{author}{\bibfnamefont{S.~R.} \bibnamefont{Nagel}},
  \bibinfo{journal}{Phys. Rev. Lett.}
  \textbf{\bibinfo{volume}{71}}(\bibinfo{number}{9}), \bibinfo{pages}{1470}
  (\bibinfo{year}{1993}).

\bibitem{sun94}
\bibinfo{author}{\bibfnamefont{T.}~\bibnamefont{Sun}},
  \bibinfo{author}{\bibfnamefont{P.}~\bibnamefont{Meakin}}, \bibnamefont{and}
  \bibinfo{author}{\bibfnamefont{T.}~\bibnamefont{J{\o}ssang}},
  \bibinfo{journal}{Phys. Rev. E}
  \textbf{\bibinfo{volume}{49}}(\bibinfo{number}{6}), \bibinfo{pages}{4865}
  (\bibinfo{year}{1994}).

\bibitem{sun94b}
\bibinfo{author}{\bibfnamefont{T.}~\bibnamefont{Sun}},
  \bibinfo{author}{\bibfnamefont{P.}~\bibnamefont{Meakin}}, \bibnamefont{and}
  \bibinfo{author}{\bibfnamefont{T.}~\bibnamefont{J{\o}ssang}},
  \bibinfo{journal}{Water Resour. Res.}
  \textbf{\bibinfo{volume}{30}}(\bibinfo{number}{9}), \bibinfo{pages}{2599}
  (\bibinfo{year}{1994}).

\bibitem{somfai97}
\bibinfo{author}{\bibfnamefont{E.}~\bibnamefont{Somfai}} \bibnamefont{and}
  \bibinfo{author}{\bibfnamefont{L.~M.} \bibnamefont{Sander}},
  \bibinfo{journal}{Phys. Rev. E}
  \textbf{\bibinfo{volume}{56}}(\bibinfo{number}{1}), \bibinfo{pages}{R5}
  (\bibinfo{year}{1997}).

\bibitem{rodriguez-iturbe97}
\bibinfo{author}{\bibfnamefont{I.}~\bibnamefont{Rodr\'{\i}guez-Iturbe}}
  \bibnamefont{and} \bibinfo{author}{\bibfnamefont{A.}~\bibnamefont{Rinaldo}},
  \emph{\bibinfo{title}{Fractal River Basins: Chance and Self-Organization}}
  (\bibinfo{publisher}{Cambridge University Press}, \bibinfo{address}{Great
  Britain}, \bibinfo{year}{1997}).

\bibitem{bak97}
\bibinfo{author}{\bibfnamefont{P.}~\bibnamefont{Bak}},
  \emph{\bibinfo{title}{How Nature Works: the Science of Self-Organized
  Criticality}} (\bibinfo{publisher}{Springer-Verlag}, \bibinfo{address}{New
  York}, \bibinfo{year}{1996}).

\bibitem{scheidegger67b}
\bibinfo{author}{\bibfnamefont{A.~E.} \bibnamefont{Scheidegger}},
  \bibinfo{journal}{International Association of Scientific Hydrology Bulletin}
  \textbf{\bibinfo{volume}{12}}(\bibinfo{number}{1}), \bibinfo{pages}{15}
  (\bibinfo{year}{1967}).

\bibitem{scheidegger90}
\bibinfo{author}{\bibfnamefont{A.~E.} \bibnamefont{Scheidegger}},
  \emph{\bibinfo{title}{Theoretical Geomorphology}}
  (\bibinfo{publisher}{Springer-Verlag}, \bibinfo{address}{New York},
  \bibinfo{year}{1991}), third ed.

\bibitem{strahler57}
\bibinfo{author}{\bibfnamefont{A.~N.} \bibnamefont{Strahler}},
  \bibinfo{journal}{EOS Trans. AGU}
  \textbf{\bibinfo{volume}{38}}(\bibinfo{number}{6}), \bibinfo{pages}{913}
  (\bibinfo{year}{1957}).

\bibitem{melton59}
\bibinfo{author}{\bibfnamefont{M.~A.} \bibnamefont{Melton}},
  \bibinfo{journal}{J. Geol.} \textbf{\bibinfo{volume}{67}},
  \bibinfo{pages}{345} (\bibinfo{year}{1959}).

\bibitem{gray61}
\bibinfo{author}{\bibfnamefont{D.~M.} \bibnamefont{Gray}}, \bibinfo{journal}{J.
  Geophys. Res.} \textbf{\bibinfo{volume}{66}}(\bibinfo{number}{4}),
  \bibinfo{pages}{1215} (\bibinfo{year}{1961}).

\bibitem{rigon96}
\bibinfo{author}{\bibfnamefont{R.}~\bibnamefont{Rigon}},
  \bibinfo{author}{\bibfnamefont{I.}~\bibnamefont{Rodr\'{\i}guez-Iturbe}},
  \bibinfo{author}{\bibfnamefont{A.}~\bibnamefont{Maritan}},
  \bibinfo{author}{\bibfnamefont{A.}~\bibnamefont{Giacometti}},
  \bibinfo{author}{\bibfnamefont{D.~G.} \bibnamefont{Tarboton}},
  \bibnamefont{and} \bibinfo{author}{\bibfnamefont{A.}~\bibnamefont{Rinaldo}},
  \bibinfo{journal}{Water Resour. Res.}
  \textbf{\bibinfo{volume}{32}}(\bibinfo{number}{11}), \bibinfo{pages}{3367}
  (\bibinfo{year}{1996}).

\bibitem{mueller72}
\bibinfo{author}{\bibfnamefont{J.~E.} \bibnamefont{Mueller}},
  \bibinfo{journal}{Geological Society of America Bulletin}
  \textbf{\bibinfo{volume}{83}}, \bibinfo{pages}{3471} (\bibinfo{year}{1972}).

\bibitem{mosley73}
\bibinfo{author}{\bibfnamefont{M.~P.} \bibnamefont{Mosley}} \bibnamefont{and}
  \bibinfo{author}{\bibfnamefont{R.~S.} \bibnamefont{Parker}},
  \bibinfo{journal}{Geological Society of America Bulletin}
  \textbf{\bibinfo{volume}{84}}, \bibinfo{pages}{3123} (\bibinfo{year}{1973}).

\bibitem{mueller73}
\bibinfo{author}{\bibfnamefont{J.~E.} \bibnamefont{Mueller}},
  \bibinfo{journal}{Geological Society of America Bulletin}
  \textbf{\bibinfo{volume}{84}}, \bibinfo{pages}{3127} (\bibinfo{year}{1973}).

\bibitem{labarbera90}
\bibinfo{author}{\bibfnamefont{P.}~\bibnamefont{La~Barbera}} \bibnamefont{and}
  \bibinfo{author}{\bibfnamefont{R.}~\bibnamefont{Rosso}},
  \bibinfo{journal}{Water Resour. Res.}
  \textbf{\bibinfo{volume}{26}}(\bibinfo{number}{9}), \bibinfo{pages}{2245}
  (\bibinfo{year}{1990}).

\bibitem{tokunaga66}
\bibinfo{author}{\bibfnamefont{E.}~\bibnamefont{Tokunaga}},
  \bibinfo{journal}{Geophys. Bull. Hokkaido Univ.}
  \textbf{\bibinfo{volume}{15}}, \bibinfo{pages}{1} (\bibinfo{year}{1966}).

\bibitem{tokunaga78}
\bibinfo{author}{\bibfnamefont{E.}~\bibnamefont{Tokunaga}},
  \bibinfo{journal}{Geogr. Rep., Tokyo Metrop. Univ.}
  \textbf{\bibinfo{volume}{13}}, \bibinfo{pages}{1} (\bibinfo{year}{1978}).

\bibitem{tokunaga84}
\bibinfo{author}{\bibfnamefont{E.}~\bibnamefont{Tokunaga}},
  \bibinfo{journal}{Trans. Jpn. Geomorphol. Union}
  \textbf{\bibinfo{volume}{5}}(\bibinfo{number}{2}), \bibinfo{pages}{71}
  (\bibinfo{year}{1984}).

\bibitem{peckham95}
\bibinfo{author}{\bibfnamefont{S.~D.} \bibnamefont{Peckham}},
  \bibinfo{journal}{Water Resour. Res.}
  \textbf{\bibinfo{volume}{31}}(\bibinfo{number}{4}), \bibinfo{pages}{1023}
  (\bibinfo{year}{1995}).

\bibitem{newman97}
\bibinfo{author}{\bibfnamefont{W.~I.} \bibnamefont{Newman}},
  \bibinfo{author}{\bibfnamefont{D.~L.} \bibnamefont{Turcotte}},
  \bibnamefont{and} \bibinfo{author}{\bibfnamefont{A.~M.}
  \bibnamefont{Gabrielov}}, \bibinfo{journal}{Fractals}
  \textbf{\bibinfo{volume}{5}}(\bibinfo{number}{4}), \bibinfo{pages}{603}
  (\bibinfo{year}{1997}).

\bibitem{scheidegger68c}
\bibinfo{author}{\bibfnamefont{A.~E.} \bibnamefont{Scheidegger}},
  \bibinfo{journal}{Water Resour. Res.}
  \textbf{\bibinfo{volume}{4}}(\bibinfo{number}{5}), \bibinfo{pages}{1015}
  (\bibinfo{year}{1968}).

\bibitem{kirchner93}
\bibinfo{author}{\bibfnamefont{J.~W.} \bibnamefont{Kirchner}},
  \bibinfo{journal}{Geology} \textbf{\bibinfo{volume}{21}},
  \bibinfo{pages}{591} (\bibinfo{year}{1993}).

\bibitem{schumm56a}
\bibinfo{author}{\bibfnamefont{S.~A.} \bibnamefont{Schumm}},
  \bibinfo{journal}{Bull. Geol. Soc. Am} \textbf{\bibinfo{volume}{67}},
  \bibinfo{pages}{597} (\bibinfo{year}{1956}).

\bibitem{takayasu88}
\bibinfo{author}{\bibfnamefont{H.}~\bibnamefont{Takayasu}},
  \bibinfo{author}{\bibfnamefont{I.}~\bibnamefont{Nishikawa}},
  \bibnamefont{and} \bibinfo{author}{\bibfnamefont{H.}~\bibnamefont{Tasaki}},
  \bibinfo{journal}{Phys. Rev. A}
  \textbf{\bibinfo{volume}{37}}(\bibinfo{number}{8}), \bibinfo{pages}{3110}
  (\bibinfo{year}{1988}).

\bibitem{takayasu89a}
\bibinfo{author}{\bibfnamefont{M.}~\bibnamefont{Takayasu}} \bibnamefont{and}
  \bibinfo{author}{\bibfnamefont{H.}~\bibnamefont{Takayasu}},
  \bibinfo{journal}{Phys. Rev. A}
  \textbf{\bibinfo{volume}{39}}(\bibinfo{number}{8}), \bibinfo{pages}{4345}
  (\bibinfo{year}{1989}).

\bibitem{takayasu89b}
\bibinfo{author}{\bibfnamefont{H.}~\bibnamefont{Takayasu}},
  \bibinfo{journal}{Physcial Review Letters}
  \textbf{\bibinfo{volume}{63}}(\bibinfo{number}{23}), \bibinfo{pages}{2563}
  (\bibinfo{year}{1989}).

\bibitem{takayasu90}
\bibinfo{author}{\bibfnamefont{H.}~\bibnamefont{Takayasu}},
  \emph{\bibinfo{title}{Fractals in the Physical Sciences}}
  (\bibinfo{publisher}{Manchester University Press},
  \bibinfo{address}{Manchester}, \bibinfo{year}{1990}).

\bibitem{takayasu91}
\bibinfo{author}{\bibfnamefont{H.}~\bibnamefont{Takayasu}},
  \bibinfo{author}{\bibfnamefont{M.}~\bibnamefont{Takayasu}},
  \bibinfo{author}{\bibfnamefont{A.}~\bibnamefont{Provata}}, \bibnamefont{and}
  \bibinfo{author}{\bibfnamefont{G.}~\bibnamefont{Huber}}, \bibinfo{journal}{J.
  Stat. Phys.} \textbf{\bibinfo{volume}{65}}(\bibinfo{number}{3/4}),
  \bibinfo{pages}{725} (\bibinfo{year}{1991}).

\bibitem{huber91}
\bibinfo{author}{\bibfnamefont{G.}~\bibnamefont{Huber}},
  \bibinfo{journal}{Physica A} \textbf{\bibinfo{volume}{170}},
  \bibinfo{pages}{463} (\bibinfo{year}{1991}).

\bibitem{abrahams84}
\bibinfo{author}{\bibfnamefont{A.~D.} \bibnamefont{Abrahams}},
  \bibinfo{journal}{Water Resour. Res.}
  \textbf{\bibinfo{volume}{20}}(\bibinfo{number}{2}), \bibinfo{pages}{161}
  (\bibinfo{year}{1984}).

\bibitem{montgomery92}
\bibinfo{author}{\bibfnamefont{D.~R.} \bibnamefont{Montgomery}}
  \bibnamefont{and} \bibinfo{author}{\bibfnamefont{W.~E.}
  \bibnamefont{Dietrich}}, \bibinfo{journal}{Science}
  \textbf{\bibinfo{volume}{255}}, \bibinfo{pages}{826} (\bibinfo{year}{1992}).

\bibitem{tarboton89}
\bibinfo{author}{\bibfnamefont{D.~G.} \bibnamefont{Tarboton}},
  \bibinfo{author}{\bibfnamefont{R.~L.} \bibnamefont{Bras}}, \bibnamefont{and}
  \bibinfo{author}{\bibfnamefont{I.}~\bibnamefont{Rodr\'{\i}guez-Iturbe}},
  \bibinfo{journal}{Water Resour. Res.}
  \textbf{\bibinfo{volume}{25}}(\bibinfo{number}{9}), \bibinfo{pages}{2037}
  (\bibinfo{year}{1989}).

\bibitem{shreve67}
\bibinfo{author}{\bibfnamefont{R.~L.} \bibnamefont{Shreve}},
  \bibinfo{journal}{J. Geol.} \textbf{\bibinfo{volume}{75}},
  \bibinfo{pages}{178} (\bibinfo{year}{1967}).

\bibitem{haggett69}
\bibinfo{author}{\bibfnamefont{P.}~\bibnamefont{Haggett}} \bibnamefont{and}
  \bibinfo{author}{\bibfnamefont{R.~J.} \bibnamefont{Chorley}},
  \emph{\bibinfo{title}{Network Analysis in Geography}}
  (\bibinfo{publisher}{Edward Arnold}, \bibinfo{address}{London},
  \bibinfo{year}{1969}).

\bibitem{gardiner73}
\bibinfo{author}{\bibfnamefont{V.}~\bibnamefont{Gardiner}},
  \bibinfo{journal}{Geogr. Ann.}
  \textbf{\bibinfo{volume}{55A}}(\bibinfo{number}{3--4}), \bibinfo{pages}{147}
  (\bibinfo{year}{1973}).

\bibitem{morisawa62}
\bibinfo{author}{\bibfnamefont{M.~E.} \bibnamefont{Morisawa}},
  \bibinfo{journal}{Geological Society of America Bulletin}
  \textbf{\bibinfo{volume}{73}}, \bibinfo{pages}{1025} (\bibinfo{year}{1962}).

\bibitem{devries94}
\bibinfo{author}{\bibfnamefont{H.}~\bibnamefont{de~Vries}},
  \bibinfo{author}{\bibfnamefont{T.}~\bibnamefont{Becker}}, \bibnamefont{and}
  \bibinfo{author}{\bibfnamefont{B.}~\bibnamefont{Eckhardt}},
  \bibinfo{journal}{Water Resour. Res.}
  \textbf{\bibinfo{volume}{30}}(\bibinfo{number}{12}), \bibinfo{pages}{3541}
  (\bibinfo{year}{1994}).

\bibitem{glock31}
\bibinfo{author}{\bibfnamefont{W.~S.} \bibnamefont{Glock}},
  \bibinfo{journal}{The Geogr. Rev.} \textbf{\bibinfo{volume}{21}},
  \bibinfo{pages}{475} (\bibinfo{year}{1931}).

\bibitem{dodds98d}
\bibinfo{author}{\bibfnamefont{P.~S.} \bibnamefont{Dodds}},
  \bibinfo{author}{\bibfnamefont{R.}~\bibnamefont{Pastor-Satorras}},
  \bibnamefont{and} \bibinfo{author}{\bibfnamefont{D.~H.}
  \bibnamefont{Rothman}}, \emph{\bibinfo{title}{Fluctuation in River network
  scaling laws}} (\bibinfo{year}{1999}), \bibinfo{note}{in preparation}.

\bibitem{rosso91}
\bibinfo{author}{\bibfnamefont{R.}~\bibnamefont{Rosso}},
  \bibinfo{author}{\bibfnamefont{B.}~\bibnamefont{Bacchi}}, \bibnamefont{and}
  \bibinfo{author}{\bibfnamefont{P.}~\bibnamefont{La~Barbera}},
  \bibinfo{journal}{Water Resour. Res.}
  \textbf{\bibinfo{volume}{27}}(\bibinfo{number}{3}), \bibinfo{pages}{381}
  (\bibinfo{year}{1991}).

\bibitem{smart67}
\bibinfo{author}{\bibfnamefont{J.~S.} \bibnamefont{Smart}},
  \bibinfo{journal}{Water Resour. Res.}
  \textbf{\bibinfo{volume}{3}}(\bibinfo{number}{3}), \bibinfo{pages}{773}
  (\bibinfo{year}{1967}).

\bibitem{feder88}
\bibinfo{author}{\bibfnamefont{J.}~\bibnamefont{Feder}},
  \emph{\bibinfo{title}{Fractals}} (\bibinfo{publisher}{Plenum Press},
  \bibinfo{address}{New York}, \bibinfo{year}{1988}).

\bibitem{labarbera94}
\bibinfo{author}{\bibfnamefont{P.}~\bibnamefont{La~Barbera}} \bibnamefont{and}
  \bibinfo{author}{\bibfnamefont{G.}~\bibnamefont{Roth}},
  \bibinfo{journal}{Hydrol. Processes} \textbf{\bibinfo{volume}{8}},
  \bibinfo{pages}{125} (\bibinfo{year}{1994}).

\bibitem{stark97}
\bibinfo{author}{\bibfnamefont{C.~P.} \bibnamefont{Stark}},
  \emph{\bibinfo{title}{Stream networks on a Bethe lattice: Cayley trees,
  invasion percolation and branching ratios}} (\bibinfo{year}{1997}),
  \bibinfo{note}{preprint}.

\bibitem{colaiori97}
\bibinfo{author}{\bibfnamefont{F.}~\bibnamefont{Colaiori}},
  \bibinfo{author}{\bibfnamefont{A.}~\bibnamefont{Flammini}},
  \bibinfo{author}{\bibfnamefont{A.}~\bibnamefont{Maritan}}, \bibnamefont{and}
  \bibinfo{author}{\bibfnamefont{J.~R.} \bibnamefont{Banavar}},
  \bibinfo{journal}{Phys. Rev. E}
  \textbf{\bibinfo{volume}{55}}(\bibinfo{number}{2}), \bibinfo{pages}{1298}
  (\bibinfo{year}{1997}).

\bibitem{cieplak98a}
\bibinfo{author}{\bibfnamefont{M.}~\bibnamefont{Cieplak}},
  \bibinfo{author}{\bibfnamefont{A.}~\bibnamefont{Giacometti}},
  \bibinfo{author}{\bibfnamefont{A.}~\bibnamefont{Maritan}},
  \bibinfo{author}{\bibfnamefont{A.}~\bibnamefont{Rinaldo}},
  \bibinfo{author}{\bibfnamefont{I.}~\bibnamefont{Rodr\'{\i}guez-Iturbe}},
  \bibnamefont{and} \bibinfo{author}{\bibfnamefont{J.~R.}
  \bibnamefont{Banavar}}, \bibinfo{journal}{J. Stat. Phys.}
  \textbf{\bibinfo{volume}{91}}(\bibinfo{number}{1/2}), \bibinfo{pages}{1}
  (\bibinfo{year}{1998}).

\bibitem{west97}
\bibinfo{author}{\bibfnamefont{G.~B.} \bibnamefont{West}},
  \bibinfo{author}{\bibfnamefont{J.~H.} \bibnamefont{Brown}}, \bibnamefont{and}
  \bibinfo{author}{\bibfnamefont{B.~J.} \bibnamefont{Enquist}},
  \bibinfo{journal}{Science} \textbf{\bibinfo{volume}{276}},
  \bibinfo{pages}{122} (\bibinfo{year}{1997}).

\end{thebibliography}
\end{document}